\documentclass{article}

\usepackage{arxiv}
\usepackage{lmodern}
\usepackage[utf8]{inputenc}
\usepackage[T1]{fontenc}
\usepackage{longtable}
\usepackage{enumitem}
\usepackage{appendix}
\usepackage{booktabs}
\usepackage{multirow}
\usepackage{graphicx}
\usepackage{float}
\usepackage{mathtools}
\usepackage[authoryear]{natbib}
\usepackage{color}
\usepackage{amsmath}
\usepackage{amsthm}
\usepackage[linesnumbered,ruled]{algorithm2e}
\usepackage{bm}
\usepackage{newtxtext}
\usepackage{newtxmath}
\usepackage{microtype}
\usepackage{url}
\usepackage{hyperref}
\hypersetup{hidelinks}
\usepackage{doi}
\usepackage{cleveref}

\allowdisplaybreaks[4]
\bibpunct[, ]{(}{)}{,}{a}{}{,}
\newtheorem{proposition}{Proposition}
\newtheorem{definition}{Definition}

\newtheorem{theorem}{Theorem}
\newtheorem{lemma}{Lemma}
\newtheorem{corollary}{Corollary}

\title{Decision Timescales in Max-Pressure Traffic Signal Control: Theory, Methodology, and Validation}

\author{%
Chaopeng Tan*\\
\normalfont Chair of Traffic Process Automation, Technische Universität Dresden\\
\normalfont Hettnerstra{\ss}e 3, 01069 Dresden, Germany\\
\normalfont \texttt{chaopeng.tan@tu-dresden.de}
}
\date{}

\renewcommand{\shorttitle}{Decision Timescales in Max-Pressure Traffic Signal Control}

\hypersetup{
  pdftitle={Decision Timescales in Max-Pressure Traffic Signal Control: Theory, Methodology, and Validation},
  pdfauthor={Chaopeng Tan},
  pdfkeywords={Max-Pressure control, decision timescale, traffic signal control, spatiotemporal heterogeneity, network stability}
}

\begin{document}
\maketitle

\begin{abstract}
Max-pressure (MP) control adapts signal phases to prevailing traffic states, yet the timescale at which these decisions are reconsidered is typically fixed a priori. This paper treats the decision timescale as an explicit dimension of MP control. We first characterize the trade-off between responsiveness and switching efficiency by quantifying how an equivalent switching loss reduces the service capacity guaranteed over a finite decision interval. This yields a load-dependent critical timescale determined by the switching loss and normalized traffic load, together with a performance-oriented operating range above the stability boundary. Building on this result, we propose Variable-Timescale Max-Pressure (VT-MP), in which each intersection selects its decision interval according to its local operating condition, allowing the control timescale to vary across both time and space. We establish strong stability of the resulting asynchronous heterogeneous controller under uniformly admissible local loads and bounded decision intervals, and introduce a downstream receiving-supply correction for finite-storage implementation. The framework is evaluated through three experiments with complementary roles. A single-intersection mechanism experiment first provides empirical validation of the theoretical load--timescale relationship, showing that the preferred decision interval increases systematically with traffic load and follows the derived characteristic timescales. Network-level effectiveness is then evaluated in a store-and-forward network consistent with the analytical assumptions and in a microscopic $5\times5$ SUMO network with finite storage. Heterogeneous VT-MP reduces total queueing time by 31.0\% relative to the best fixed decision interval benchmark in the store-and-forward network. In the microscopic SUMO experiment, it reduces mean trip delay by 24.3\%, from 213.3~s to 161.5~s, and peak network accumulation by 12.3\% relative to the best fixed decision interval benchmark. These findings indicate that the MP decision timescale should be treated as a traffic-state-dependent control variable rather than a fixed implementation parameter.
\end{abstract}


\section{Introduction}

Urban traffic signal control requires allocating limited intersection capacity in response to traffic conditions that vary across both space and time, a problem that has motivated a broad range of traffic-responsive and network-wide control strategies \citep{papageorgiou2003review,diakaki2002tuc,aboudolas2009store}. Among network-level traffic-responsive control approaches, Max-Pressure (MP) control has attracted sustained interest because it combines a decentralized control structure with rigorous network-level stability properties. Originating from backpressure scheduling for constrained queueing networks \citep{tassiulas1992stability}, MP selects signal phases according to pressure differences between upstream and downstream traffic states. Its application to urban traffic networks established that locally computed signal decisions can stabilize network queues for admissible demand under standard store-and-forward assumptions \citep{wongpiromsarn2012distributed,varaiya2013max}. Subsequent work further developed decentralized MP formulations and examined their implementation in urban networks \citep{le2015decentralized}. The resulting combination of traffic responsiveness, low computational complexity, decentralized operation, and analytical stability guarantees has made MP an important framework for network traffic signal control \citep{levin2023review}.

A major direction in the development of MP has been to enrich the traffic information represented by pressure. The original queue-based formulation, commonly referred to as Q-MP, uses vehicle queues on upstream and downstream movements to characterize local traffic imbalance \citep{varaiya2013max}. Capacity-aware backpressure modifies this representation to account for finite downstream storage and congestion propagation \citep{gregoire2015capacity}, while position-weighted backpressure (PW-MP) incorporates the spatial distribution of vehicles along links \citep{li2019position}. Other studies have explored traffic states that can be obtained without relying exclusively on conventional queue measurements. Travel-time-based pressure uses observed travel times to characterize traffic conditions \citep{mercader2020travel}, crowdsourced delay information has been used for pressure-based traffic signal control \citep{dixit2020crowdsourced}, and Delay-based Max-Pressure (D-MP) formally uses vehicle delay as the pressure metric while retaining the maximum-stability property \citep{liu2022delay}. More recently, connected-vehicle information has enabled traffic state estimation under incomplete and spatially heterogeneous observations \citep{tan2021queue,tan2022cfd,islam2020partial,comert2021queue,tan2026privatemp}. CV-MP, for example, constructs pressure from connected-vehicle travel-time information and establishes stability conditions under partial and heterogeneous connectivity \citep{tan2026cvmp}. Collectively, these developments have progressively expanded the traffic states available to MP, from aggregate queues to spatial, temporal, and connected-vehicle observations.

A parallel research direction has extended MP to a broader range of traffic-management objectives and operating environments. Practical signal operation has been incorporated through cycle-based and cyclical MP formulations \citep{anderson2018cycle,levin2020cyclical}. For public transport, MP has been integrated with transit signal priority for bus rapid transit operations \citep{xu2022transit}; Priority-MP further combines MP-based signal control with dwelling and speed advisory mechanisms for transit priority \citep{tan2026priority}; and Transit-MP extends transit-prioritized MP to sparse connected-vehicle environments \citep{tan2026transit}. MP has also been extended to explicitly incorporate pedestrian queues while maintaining decentralized control and network-stability properties \citep{liu2024pedestrian}. At a larger spatial scale, MP has been coupled with macroscopic perimeter control \citep{tsitsokas2023twolayer} and subsequently extended to incorporate regional network states directly into the pressure framework through N-MP \citep{liu2024nmp}. These studies demonstrate how the basic pressure-control principle can accommodate increasingly diverse modes, network states, and traffic-management objectives while preserving much of its decentralized structure. Related adaptive-signal research has likewise emphasized decentralized response and network coordination, ranging from self-organizing signal rules to multi-agent reinforcement learning \citep{gershenson2005self,arel2010rl,eltantawy2013marlin,rasheed2020review}. More recently, learning-based traffic signal control has incorporated transportation-domain structure at larger scales, including pressure-informed reward design in PressLight and network-level multi-agent coordination in CoLight and large-scale decentralized control \citep{wei2019presslight,wei2019colight,chen2020thousand}.

Alongside the choice of traffic state and control application, MP contains another fundamental design dimension: the timescale at which pressure is re-evaluated and a signal decision is reconsidered. In conventional time-step implementations, pressure is evaluated at a prescribed control interval, whereas cycle-based formulations organize MP decisions within a predefined or bounded cycle structure \citep{anderson2018cycle,levin2020cyclical}. The selected timescale determines how long a pressure-based decision remains in effect before the controller reacts to newly evolving traffic conditions. Thus, MP control involves not only determining \emph{which} phase should receive service, but also determining \emph{when} that decision should be reconsidered. The latter is commonly specified through a control step, decision interval, cycle, or holding period rather than derived directly from the prevailing traffic load. This motivates a basic question for MP control: \emph{what is an appropriate decision timescale for a given traffic operating condition?}

The importance of this question arises from a fundamental trade-off between control responsiveness and switching efficiency. Short decision intervals allow pressure to be updated frequently and enable the controller to react rapidly to evolving queue imbalances. However, frequent changes of signal phase increase the fraction of time consumed by yellow, all-red, startup, and other transition losses. Conversely, longer intervals reduce the relative cost of switching but retain each pressure decision for a longer period, during which the traffic state may evolve and the selected phase may become progressively less representative of current conditions. Previous studies have incorporated switching losses explicitly into MP control. Switching-Curve Max-Pressure, for example, introduces a switching rule that accounts for phase-switching loss while retaining throughput-optimality \citep{wang2022switching}, and cycle-constrained MP formulations regulate phase changes through structured signal operation \citep{anderson2018cycle,levin2020cyclical}. These developments establish the importance of signal-transition dynamics. At the same time, they raise a complementary timescale question: how should the duration between successive MP decisions vary with the traffic load itself?

This question becomes particularly relevant because the operating condition of an urban network is neither temporally nor spatially uniform. During low-demand periods, traffic states can change substantially relative to the small number of queued vehicles, favoring rapid response, whereas under heavy demand the service lost to repeated phase switching becomes increasingly consequential. The appropriate balance can therefore change as demand evolves over time. Moreover, different intersections within the same network may simultaneously operate at very different levels of saturation. A single network-wide decision interval, even if adjusted over time, imposes the same control timescale on lightly loaded and heavily loaded intersections. This suggests a more general formulation in which the MP decision timescale is allowed to vary by both node and time, so that each intersection operates at a timescale consistent with its local traffic condition.

Motivated by this observation, this study develops \emph{Variable-Timescale Max-Pressure} (VT-MP) control. We first analyze the role of the decision interval within a store-and-forward MP network with nonzero switching loss. The analysis shows that switching time directly reduces the effective service available within a decision frame and yields a load-dependent critical decision timescale. A performance bound then characterizes the trade-off between switching efficiency and state responsiveness and identifies a performance-oriented operating range above the stability boundary. Building on these results, VT-MP allows each intersection to select a local, time-varying decision interval according to its operating condition. Consequently, different intersections may update asynchronously while retaining the decentralized structure of MP. We further establish network stability for the resulting spatiotemporally heterogeneous controller under uniformly admissible local loads and bounded decision intervals. For microscopic networks with finite storage, the implementation is augmented with a local downstream receiving-supply constraint.

The main contributions of this study are threefold.
\begin{enumerate}
    \item We characterize the effect of signal-switching loss on the service available under a finite MP decision interval. This yields an explicit switching-adjusted capacity guarantee, the load-dependent critical timescale $T_c=I/(1-\rho)$, and a performance-oriented range $T_c<T\le2T_c$ obtained from a finite-frame queue bound. The analysis, for the first time, provides a direct theoretical connection between the temporal resolution of MP decisions, traffic load, and effective switching loss.

    \item We develop VT-MP by making the decision timescale a local control variable rather than a network-wide fixed setting. The resulting controller permits node-specific and time-varying decision intervals and therefore asynchronous MP updates across the network. We establish strong stability under uniformly admissible local loads and bounded local intervals, extend the result to sufficiently persistent piecewise-stationary demand, and show analytically how heterogeneous timing can improve realized service utilization while relaxing the performance restriction imposed by a common network-wide timescale.

    \item We first validate the theoretical load--timescale mechanism and then evaluate the network-level effectiveness of VT-MP. A single-intersection mechanism experiment confirms that the preferred MP decision timescale increases with traffic load and follows the derived characteristic scales; it also provides the timing-factor calibration used in the network experiments. In the store-and-forward network, heterogeneous VT-MP reduces total queueing time by 31.0\% relative to the best fixed interval. In the microscopic SUMO network with finite storage, the same controller reduces mean trip delay by 24.3\% and peak network accumulation by 12.3\% relative to the seed-specific best fixed benchmark.
\end{enumerate}

The remainder of this paper is organized as follows. Section~\ref{sec:problem} introduces the MP model and decision-timescale problem, and Section~\ref{sec:theory} develops the corresponding timescale theory. Section~\ref{sec:vtmp} presents VT-MP and its stability and implementation, while Section~\ref{sec:experiments} validates the theory and evaluates network-level performance. Section~\ref{sec:conclusion} concludes the paper.

\section{Max-Pressure Control and the Decision-Timescale Problem}
\label{sec:problem}

This section establishes the conventional Max-Pressure (MP) framework on which the subsequent decision-timescale analysis is built. We first describe the movement-level traffic dynamics and then introduce the MP phase-selection rule together with the classical capacity, admissibility, and stability concepts associated with it. These established results provide a common analytical baseline rather than a new proof contribution. We then introduce finite decision frames and switching loss and formulate the additional temporal question studied in this paper: how long an MP phase decision should be retained before pressure is reconsidered.

\subsection{Network model and conventional Max-Pressure control}
\label{subsec:network_model}

This subsection defines the baseline store-and-forward network and conventional MP controller. The traffic dynamics specify the movement demand that must be served, while the MP rule determines how the available signal service is allocated. We then summarize the classical capacity and stability concepts needed later to distinguish the effect of changing the decision timescale from the established spatial phase-allocation mechanism of MP.

\subsubsection{Traffic dynamics}

Consider a directed road network represented by a graph $\mathcal{G}=(\mathcal{V},\mathcal{E})$, where $\mathcal{V}$ is the set of signalized intersections and $\mathcal{E}$ is the set of directed links. For two consecutive links $a,b\in\mathcal{E}$, the ordered pair $(a,b)$ denotes a feasible movement from link $a$ to link $b$, and $\mathcal{M}$ denotes the set of all feasible movements. Let $\mathcal{M}_n\subseteq\mathcal{M}$ be the set of movements controlled by intersection $n$.

Let $Q_{ab}(t)$ denote the number of vehicles at time $t$ that are on link $a$ and intend to enter link $b$. For a movement $(b,c)$, let $r_{bc}$ denote the fraction of traffic entering link $b$ that subsequently turns toward link $c$. The turning proportions satisfy $r_{bc}\ge0$ and $\sum_{c:(b,c)\in\mathcal{M}}r_{bc}\le1$; the remaining fraction leaves the modeled network after traversing link $b$. This movement-based state is convenient for MP because it preserves both the queue requesting service and the downstream direction into which the discharged vehicles will be routed.

Each signalized intersection $n\in\mathcal V$ has a finite set of feasible phases $\mathcal P_n$. For phase $p\in\mathcal P_n$, let $\mu_{ab}(p)$ denote the saturation service rate assigned to movement $(a,b)$. Every modeled movement is assumed to receive positive service under at least one feasible phase. Let $A^{\mathrm{ext}}_{ab}(t)$ be the exogenous arrivals to movement $(a,b)$ during $[t,t+1)$, and let $D_{ab}(t)$ be the realized departures. The movement queues evolve as
\begin{equation}
Q_{ab}(t+1)
=
Q_{ab}(t)
-
D_{ab}(t)
+
A^{\mathrm{ext}}_{ab}(t)
+
r_{ab}\sum_{c:(c,a)\in\mathcal M}D_{ca}(t).
\label{eq:queue_dynamics}
\end{equation}
The first two terms describe the vehicles already associated with movement $(a,b)$, while the final term transfers departures from upstream movements into $(a,b)$ according to the turning proportions. For a productive slot in which phase $p_n(t)$ is active, define the potential service as $S_{ab}(t)=\mu_{ab}(p_n(t))$. The realized departure is
\begin{equation}
D_{ab}(t)=\min\{Q_{ab}(t),S_{ab}(t)\}.
\label{eq:departure_constraint}
\end{equation}
During a transition-loss interval, $S_{ab}(t)=0$ for the affected movements. The distinction between potential service and realized departure will be important later: MP selects a phase according to the service that the phase can offer, while a short queue may be unable to use all of that service.

The exogenous arrivals and movement service rates are assumed to be uniformly bounded. Let $\mathbb E[A^{\mathrm{ext}}_{ab}(t)\mid\mathcal F_t]=\lambda_{ab}(t)$, where $\mathcal F_t$ is the system history up to time $t$. For a fixed mean operating condition, write $\boldsymbol\lambda(t)=\boldsymbol\lambda$. The routing matrix induced by the turning proportions is assumed to describe an open network, so vehicles eventually leave the modeled system. Under a fixed mean arrival vector, open routing induces a unique routing-consistent effective movement-flow vector $\boldsymbol\nu(\boldsymbol\lambda)$ satisfying
\begin{equation}
\nu_{ab}
=
\lambda_{ab}
+
r_{ab}\sum_{c:(c,a)\in\mathcal M}\nu_{ca},
\qquad
(a,b)\in\mathcal M.
\label{eq:effective_flow_balance}
\end{equation}
The vector $\boldsymbol\nu$ includes both external demand and traffic transferred from upstream movements and therefore represents the long-run movement flow that the signal system must support. As in the standard store-and-forward MP model, links are assumed to have sufficient storage so that downstream receiving capacity does not directly constrain Eq.~\eqref{eq:queue_dynamics}. Finite-storage effects are introduced later as an implementation layer.

\subsubsection{Max-Pressure control and classical stability}
\label{subsec:classical_mp_stability}

Following the standard movement-based MP formulation \citep{varaiya2013max,le2015decentralized}, the pressure of movement $(a,b)$ is
\begin{equation}
w_{ab}(t)
=
Q_{ab}(t)
-
\sum_{c:(b,c)\in\mathcal M}r_{bc}Q_{bc}(t).
\label{eq:movement_pressure}
\end{equation}
The pressure has a direct traffic interpretation: a large positive value indicates that serving $(a,b)$ relieves a relatively large upstream accumulation compared with the downstream queues to which the discharged vehicles are expected to travel. For a feasible phase $p\in\mathcal P_n$, its phase pressure is
\[
P_{n,p}(t)
=
\sum_{(a,b)\in\mathcal M_n}\mu_{ab}(p)w_{ab}(t).
\]
Conventional MP selects
\begin{equation}
p_n^*(t)
=
\arg\max_{p\in\mathcal P_n}P_{n,p}(t).
\label{eq:mp_phase_selection}
\end{equation}
Thus, conventional MP allocates the currently available intersection service to the feasible phase with the largest pressure-weighted benefit. This rule answers the spatial allocation question---which conflicting set of movements should receive service at the current decision epoch. Before introducing the additional question of how long that decision should be retained, we summarize the classical stability structure associated with Eq.~\eqref{eq:mp_phase_selection}.

The phase-selection rule above is accompanied by the classical maximum-stability result of MP. Because the present paper changes the temporal organization of MP decisions while retaining the same pressure-based phase ranking, it is useful to state the capacity and stability concepts that will be inherited by the later analysis. For a fixed mean exogenous demand $\boldsymbol\lambda$, Eq.~\eqref{eq:effective_flow_balance} gives the routing-consistent movement flow $\boldsymbol\nu(\boldsymbol\lambda)$ that must be served throughout the network. The nominal capacity region then describes whether feasible long-run signal allocations can support these movement flows.

\begin{definition}[Nominal capacity region and admissibility]
\label{def:capacity_region}
The mean arrival vector $\boldsymbol\lambda$ is \emph{nominally feasible} if there exist nonnegative phase fractions $\theta_{n,p}$, with $\sum_{p\in\mathcal P_n}\theta_{n,p}=1$ for every intersection $n$, such that
\[
\nu_{ab}(\boldsymbol\lambda)
\le
\sum_{p\in\mathcal P_n}\theta_{n,p}\mu_{ab}(p),
\qquad
(a,b)\in\mathcal M_n.
\]
The set of all nominally feasible exogenous arrival vectors is the nominal capacity region $\mathcal C$. An arrival vector is \emph{strictly admissible} if $\boldsymbol\lambda\in\operatorname{int}(\mathcal C)$.
\end{definition}

The definition separates the traffic that must be served from the service that the signal system can provide. Routing converts the external demand into movement flows, while the phase fractions describe the long-run service available from feasible signal phases. To express the operating point on a common scale, define the normalized network load as
\begin{equation}
\rho(\boldsymbol\lambda)
=
\inf\left\{r>0:\boldsymbol\lambda\in r\mathcal C\right\}.
\label{eq:normalized_load}
\end{equation}
We write $\rho$ when its dependence on $\boldsymbol\lambda$ is clear. Thus, $\rho<1$ corresponds to strictly admissible demand, while $1-\rho$ represents the normalized reserve between the current demand and the nominal capacity boundary.

The queue-stability notion used throughout the paper is strong stability.

\begin{definition}[Strong stability]
\label{def:strong_stability}
The movement-queue process is \emph{strongly stable} if
\begin{equation}
\limsup_{M\to\infty}
\frac{1}{M}
\sum_{t=0}^{M-1}
\mathbb E\!\left[\|\mathbf Q(t)\|_1\right]
<\infty.
\label{eq:strong_stability_definition}
\end{equation}
\end{definition}

Under the standard bounded-arrival, bounded-service, and open-routing assumptions, conventional slot-by-slot MP is known to strongly stabilize every strictly admissible mean arrival vector \citep{tassiulas1992stability,varaiya2013max,le2015decentralized}. The classical proof uses the quadratic Lyapunov function $V(\mathbf Q)=\frac12\sum_{(a,b)\in\mathcal M}Q_{ab}^2$ and follows the standard stochastic-network drift framework \citep{neely2010stochastic}, establishing a one-slot drift of the form
\begin{equation}
\mathbb E\!\left[
V(\mathbf Q(t+1))-V(\mathbf Q(t))
\mid\mathcal F_t
\right]
\le
B_{\mathrm{MP}}
-
\epsilon_{\mathrm{MP}}\|\mathbf Q(t)\|_1,
\label{eq:classical_mp_drift}
\end{equation}
for finite $B_{\mathrm{MP}}$ and some $\epsilon_{\mathrm{MP}}>0$. We take this established result as the stability baseline and do not reproduce its proof here.

Two standard relations from the classical MP analysis will be reused later. First, reindexing internally routed flows gives the flow--pressure identity
\begin{equation}
\sum_{(a,b)\in\mathcal M}Q_{ab}\lambda_{ab}
=
\sum_{(a,b)\in\mathcal M}w_{ab}\nu_{ab}.
\label{eq:flow_pressure_identity}
\end{equation}
Second, open routing implies that the positive pressure magnitude controls the queue magnitude: there exists a network-dependent constant $\gamma_R>0$ such that
\begin{equation}
\|\mathbf w^+(t)\|_1
\ge
\gamma_R\|\mathbf Q(t)\|_1.
\label{eq:pressure_queue_norm}
\end{equation}
Both relations are standard ingredients in MP stability arguments \citep{varaiya2013max,le2015decentralized}. They will be reused without re-deriving the classical slot-by-slot stability proof. The new analysis begins from this baseline and asks how the same service-reserve mechanism changes when an MP decision is held for a finite duration, incurs switching loss, and later becomes asynchronous across intersections.

\subsection{Decision frames and switching loss}
\label{subsec:switching_loss}

Conventional MP specifies which phase should receive service whenever a decision is made. A practical controller must additionally specify how long that decision remains in effect before pressure is recomputed. Let $T\ge1$ denote the number of basic time slots between two successive MP evaluations. Under a fixed-timescale implementation, a decision made at frame epoch $t_k$ is retained over $[t_k,t_k+T)$, except for any transition interval required when the selected phase differs from the preceding one. Thus, $T$ is not a green time prescribed independently of MP; it is the temporal resolution at which the pressure-based phase allocation is reconsidered.

A phase transition typically includes yellow and all-red intervals and may also include startup effects before effective discharge is restored. We summarize these effects by an equivalent switching loss $I>0$, measured in units of the basic time slot. The equivalent-loss interpretation places yellow, clearance, and startup effects on the same service-time scale. If one transition occurs at the beginning of a frame of length $T$, only $T-I$ slots are guaranteed to remain productive, giving
\begin{equation}
\phi(T)
=
\frac{T-I}{T}
=
1-\frac{I}{T},
\qquad T>I.
\label{eq:service_fraction}
\end{equation}
If consecutive decisions retain the same phase, no new transition loss is incurred. Equation~\eqref{eq:service_fraction} therefore gives a conservative service fraction obtained by reserving one equivalent loss in every decision frame.

The interval $T$ consequently affects MP in two opposing ways. With a short frame, the pressure field is refreshed frequently and the controller can react quickly when queue imbalances change, but the fixed loss $I$ occupies a larger fraction of available service time. With a long frame, the relative switching loss becomes small, but the controller continues to execute a phase chosen from an increasingly old traffic state. The appropriate decision timescale must therefore balance switching efficiency against traffic-state responsiveness.

\subsection{The decision-timescale problem}
\label{subsec:timescale_problem}

The classical stability framework gives a natural starting point for this balance. The normalized load $\rho$ identifies how much of the nominal capacity region is already occupied by the mean traffic demand, while $1-\rho$ represents the remaining normalized service reserve. Switching consumes part of this reserve: for a fixed absolute loss $I$, the fraction lost in each decision frame is $I/T$. The same switching time can therefore be accommodated more readily under light traffic but becomes increasingly important as $\rho$ approaches one.

The decision-timescale problem studied in this paper is to determine how $T$ should vary with the operating load while preserving a positive service margin and avoiding unnecessarily slow response to changing queues. In the common-timescale analysis of Section~\ref{sec:theory}, this means relating $T$ to the switching loss $I$ and the nominal reserve $1-\rho$. Section~\ref{sec:vtmp} then localizes the same principle so that intersections experiencing different operating conditions can use different decision intervals.

\section{Decision-Timescale Theory for Max-Pressure Control}
\label{sec:theory}

This section develops the decision-timescale theory as an extension of the classical MP stability framework summarized in Section~\ref{subsec:classical_mp_stability}. The pressure definition, feasible signal service, and nominal capacity region remain unchanged; the new element is that an MP decision is held for a finite frame and may lose part of that frame to a phase transition. We first determine how this lost service changes the guaranteed capacity margin and obtain a load-dependent critical timescale. We then account for the opposite effect of a long frame: the queue state continues to evolve while the selected phase is held. A finite-frame Lyapunov analysis makes this responsiveness--switching-efficiency trade-off explicit and yields a performance-oriented characteristic timescale. The final subsection converts the resulting relationship into a normalized timing rule that can later be applied locally at each intersection.

\subsection{Switching-adjusted service and the critical timescale}
\label{subsec:stability_timescale}

This subsection establishes the stability-oriented lower boundary on the decision interval. The key observation is that a finite switching loss does not change the nominal phase-service vectors themselves; instead, it reduces the fraction of each decision frame during which those service vectors can be used. We therefore first translate the lost time into a conservative inner capacity region. Comparing the demand level $\rho$ with this reduced region then reveals the minimum frame length needed to retain a positive service reserve.

\subsubsection{Switching-adjusted guaranteed inner region}

For a fixed frame length $T>I$, let $\mathcal{C}^{\mathrm{g}}_T$ denote the arrival-rate region guaranteed under the conservative convention that one equivalent switching loss $I$ is reserved in every frame.

\begin{proposition}[Switching-adjusted guaranteed inner capacity region]
\label{prop:capacity_region}
For $T>I$,
\begin{equation}
\mathcal{C}^{\mathrm{g}}_T
=
\left(1-\frac{I}{T}\right)\mathcal{C}.
\label{eq:capacity_contraction}
\end{equation}
Thus, every arrival vector in the interior of $\mathcal{C}^{\mathrm{g}}_T$ is supportable under the conservative one-loss-per-frame service allocation.
\end{proposition}

\begin{proof}
Let $\phi_T=1-I/T$ denote the guaranteed productive fraction of each decision frame. Since the nominal capacity region $\mathcal C$ is generated by feasible long-run phase allocations, restricting every such allocation to the productive fraction $\phi_T$ uniformly scales the corresponding long-run service rates by $\phi_T$. Therefore,
$\mathcal C_T^{\mathrm g}
=
\phi_T\mathcal C
=
\left(1-{I}/{T}\right)\mathcal C$.
\end{proof}

\begin{figure}
    \centering
    \includegraphics[width=0.5\linewidth]{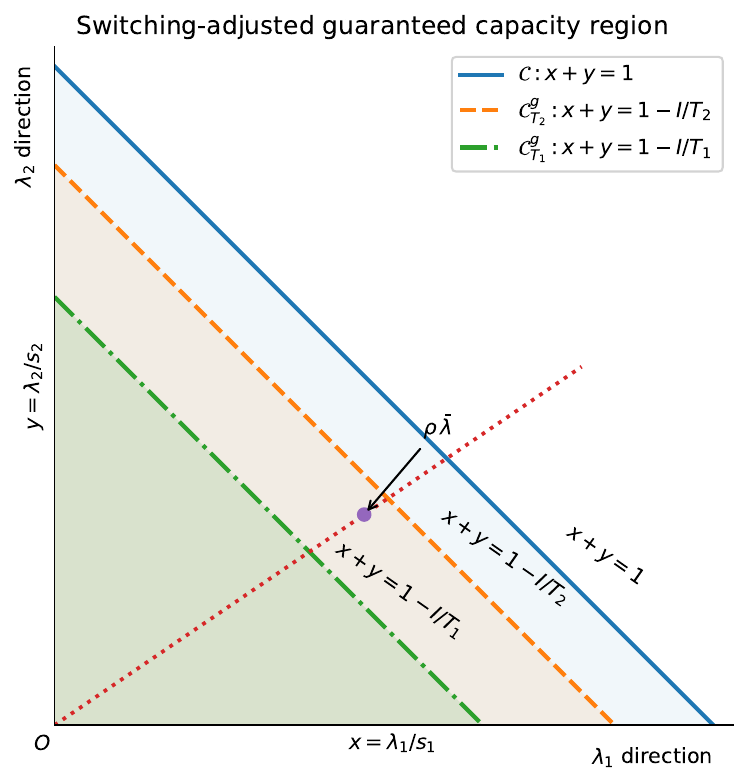}
    \caption{Switching-adjusted guaranteed capacity region in an exact two-phase normalized special case. The nominal region is $x+y\le1$, while reserving one equivalent switching loss $I$ per decision frame scales the guaranteed region to $x+y\le1-I/T$. A longer decision interval therefore retains a larger fraction of the nominal service region.}
    \label{fig:capacity_region_illustration}
\end{figure}

Figure~\ref{fig:capacity_region_illustration} gives an exact two-dimensional special case of Eq.~\eqref{eq:capacity_contraction}. Consider two mutually exclusive phases serving two normalized flow components $x=\lambda_1/s_1$ and $y=\lambda_2/s_2$. Their nominal capacity region is $\mathcal C=\{(x,y):x+y\le1,\;x,y\ge0\}$. Reserving one switching loss $I$ in each decision frame changes the boundary to $x+y\le1-I/T$. Hence, for $T_2>T_1>I$, the corresponding guaranteed regions satisfy $\mathcal C_{T_1}^{\mathrm g}\subset\mathcal C_{T_2}^{\mathrm g}\subset\mathcal C$. The marked operating point $\rho\bar{\boldsymbol\lambda}$ lies on a fixed demand ray, with $\bar{\boldsymbol\lambda}$ denoting the point where that ray intersects the nominal boundary. The illustration shows geometrically how increasing $T$ enlarges the guaranteed region available to support the same demand.

Equation~\eqref{eq:capacity_contraction} gives a guaranteed inner region rather than an exact description of the service region realized by every signal sequence. The distinction matters because two consecutive MP decisions may select the same phase, in which case the second frame does not incur a new transition loss. Reserving one loss in every frame deliberately ignores this favorable event. The resulting region is therefore conservative, but it has the advantage of depending only on $T$ and $I$, not on the future sequence of phase repetitions. This is precisely the form needed to identify a traffic-load-dependent lower boundary on $T$.

For a fixed interval $T$, we call a mean arrival vector \emph{guaranteed admissible} when $\boldsymbol\lambda\in\operatorname{int}(\mathcal C_T^{\mathrm g})$. This terminology distinguishes the conservative sufficient region used in the proof from the potentially larger realized region that may arise when consecutive decisions retain the same phase.

\subsubsection{Critical decision timescale}

Since $\boldsymbol{\lambda}\in\rho\mathcal C$ by Eq.~\eqref{eq:normalized_load}, the demand remains strictly inside the guaranteed region whenever $\rho<1-I/T$. This comparison has a simple interpretation: the nominal unused fraction $1-\rho$ must be larger than the fraction $I/T$ of each frame that is lost to switching.

\begin{proposition}[Critical decision timescale]
\label{prop:critical_timescale}
For $0\le\rho<1$ and $I>0$, a sufficient condition for the arrival vector to lie strictly inside $\mathcal{C}^{\mathrm{g}}_T$ is
\begin{equation}
T>T_c(\rho),
\qquad
T_c(\rho)
=
\frac{I}{1-\rho}.
\label{eq:critical_timescale}
\end{equation}
\end{proposition}

\begin{proof}
The condition $\rho<1-I/T$ is equivalent to $(1-\rho)T>I$, and because $1-\rho>0$, this is equivalent to $T>I/(1-\rho)$, which gives Eq.~\eqref{eq:critical_timescale}.
\end{proof}

The quantity $T_c$ is a stability-oriented boundary rather than a preferred operating interval. It is the frame length at which the normalized reserve $1-\rho$ is exactly consumed by the switching-loss fraction $I/T$. Since $\mathrm dT_c/\mathrm d\rho=I/(1-\rho)^2>0$, the boundary increases monotonically with traffic load and diverges as $\rho\rightarrow1^-$. Under light traffic, the network has sufficient spare service for more frequent phase reconsideration. Near capacity, the same switching time consumes a much larger share of the remaining reserve, so the frame must be longer simply to maintain a positive conservative service margin. What $T_c$ does not answer is whether operating just above this boundary gives good queue performance; that requires accounting for the opposite cost of holding an MP decision for a long frame.

\subsection{Performance-oriented decision timescale}
\label{subsec:performance_timescale}

Proposition~\ref{prop:critical_timescale} identifies when the conservative service margin becomes positive, but it does not describe how queue performance varies within the stable region. A very long frame makes switching relatively inexpensive, yet the selected phase is then held while the queue state continues to change. To expose this second effect, consider the standard quadratic Lyapunov function $V(\mathbf Q)=\frac12\sum_{(a,b)\in\mathcal M}Q_{ab}^2$ and examine its drift over an entire decision frame of length $T$. The resulting bound separates a frame-length penalty from the stabilizing service reserve, which makes the role of the decision timescale explicit.

\begin{lemma}[Finite-frame drift with switching loss]
\label{lem:Tslot_drift}
Suppose that arrivals and service rates are uniformly bounded, the mean arrival vector has normalized load $\rho<1$, and a phase selected at a frame epoch is retained throughout the productive portion of that frame. For any $T>T_c(\rho)$, there exist constants $B>0$ and $\chi>0$, independent of $T$ and $\mathbf{Q}(t)$, such that
\begin{equation}
\mathbb{E}
\left[
V(\mathbf{Q}(t+T))-V(\mathbf{Q}(t))
\mid \mathcal{F}_t
\right]
\le
BT^2
-
\chi\big[(1-\rho)T-I\big]\|\mathbf{Q}(t)\|_1.
\label{eq:Tslot_drift}
\end{equation}
\end{lemma}

\begin{proof}
Applying the standard quadratic queue inequality to Eq.~\eqref{eq:queue_dynamics}, reindexing the routed-flow terms, and using Eq.~\eqref{eq:movement_pressure} give the one-slot bound
\begin{equation}
\mathbb{E}
\left[
V(\mathbf{Q}(t+1))-V(\mathbf{Q}(t))
\mid \mathcal{F}_t
\right]
\le
B_0
+
\sum_{(a,b)\in\mathcal{M}}
Q_{ab}(t)\lambda_{ab}
-
\mathbb{E}
\left[
\sum_{(a,b)\in\mathcal{M}}
w_{ab}(t)D_{ab}(t)
\mid \mathcal{F}_t
\right]
\label{eq:one_slot_drift_main}
\end{equation}
for some finite $B_0$.

Let $U_{ab}(t)=S_{ab}(t)-D_{ab}(t)\ge0$ denote unused potential service. The realized-service term can be decomposed as
\[
-\sum_{(a,b)} w_{ab}D_{ab}
=
-\sum_{(a,b)} w_{ab}S_{ab}
+
\sum_{(a,b)} w_{ab}U_{ab}.
\]
Because $U_{ab}(t)>0$ only when $Q_{ab}(t)<S_{ab}(t)$, the positive part of the unused-service correction is uniformly bounded. Hence, for some $B_u<\infty$,
\[
-\sum_{(a,b)}w_{ab}(t)D_{ab}(t)
\le
-\sum_{(a,b)}w_{ab}(t)S_{ab}(t)+B_u.
\]
This converts the realized-departure term in the queue drift to the potential-service term that is maximized by MP.

Over a frame, write
\[
\mathbf Q(t+\tau)
=
\mathbf Q(t)+[\mathbf Q(t+\tau)-\mathbf Q(t)],
\qquad
\|\mathbf Q(t+\tau)-\mathbf Q(t)\|_1\le G\tau,
\]
where the second relation follows from uniformly bounded one-slot increments. Because pressure is a linear function of the queue vector, the pressure mismatch created by continuing to execute a phase selected from $\mathbf Q(t)$ also grows at most linearly with $\tau$. Accumulating this mismatch over $\tau=0,\ldots,T-1$ therefore produces an $O(T^2)$ term. This is the mathematical expression of the responsiveness cost of a long decision frame.

It remains to bound the frame-initial term. The conservative productive fraction is $\phi(T)=1-I/T$, so the radial reserve relative to the nominal load is $\phi(T)-\rho>0$. The standard MP comparison argument first gives a pressure-domain bound: for some network-dependent constant $\chi_r>0$,
\begin{align}
&\mathbb{E}\Bigg[
\sum_{\tau=0}^{T-1}
\Bigg(
\sum_{(a,b)\in\mathcal{M}}
Q_{ab}(t)A^{\mathrm{ext}}_{ab}(t+\tau)
-
\sum_{(a,b)\in\mathcal{M}}
w_{ab}(t)S_{ab}(t+\tau)
\Bigg)
\;\Bigg|\;
\mathcal{F}_t
\Bigg]
\le
-\chi_rT\big[\phi(T)-\rho\big]
\|\mathbf{w}^{+}(t)\|_1,
\label{eq:pressure_reserve_bound}
\end{align}
where $\mathbf{w}^{+}$ is the componentwise positive part of the movement-pressure vector. The open-routing relation established for conventional MP in Eq.~\eqref{eq:pressure_queue_norm} gives $\|\mathbf w^+(t)\|_1\ge\gamma_R\|\mathbf Q(t)\|_1$ for some $\gamma_R>0$. Combining this relation with Eq.~\eqref{eq:pressure_reserve_bound}, and defining $\chi=\chi_r\gamma_R$, converts the pressure-domain reserve into the queue-domain term $-\chi[(1-\rho)T-I]\|\mathbf Q(t)\|_1$. Combining this term with Eq.~\eqref{eq:one_slot_drift_main}, the bounded unused-service correction, and the $O(T^2)$ state-evolution term yields Eq.~\eqref{eq:Tslot_drift}. The full derivation and explicit constant bounds are provided in Appendix~\ref{app:Tslot_drift}.
\end{proof}

Lemma~\ref{lem:Tslot_drift} extends the classical slot-by-slot MP stability argument to finite decision frames with an explicit switching loss. In place of the nominal service reserve $1-\rho$, the effective reserve becomes $(1-\rho)T-I$ over a frame of length $T$. Hence, once this switching-adjusted reserve is positive, the negative-drift mechanism underlying classical MP is retained, yielding the following stability result.

\begin{corollary}[Strong stability of fixed-timescale MP with switching loss]
\label{cor:fixedT_stability}
Under the assumptions of Lemma~\ref{lem:Tslot_drift}, if
\[
\rho<1-\frac{I}{T},
\qquad\text{equivalently}\qquad
T>T_c(\rho),
\]
then the fixed-timescale MP controller has a finite long-run all-slot average expected total movement queue,
\[
\limsup_{M\to\infty}
\frac{1}{M}
\sum_{t=0}^{M-1}
\mathbb E\!\left[\|\mathbf Q(t)\|_1\right]
<\infty.
\]
Thus, guaranteed admissibility for interval $T$ is a sufficient condition for strong stability in the sense of Definition~\ref{def:strong_stability}.
\end{corollary}

\begin{proof}
Under the stated condition, the frame slack $\Delta_T(\rho)=(1-\rho)T-I$ is strictly positive. Lemma~\ref{lem:Tslot_drift} therefore provides a $T$-step drift of the form
\[
\mathbb E\!\left[V(\mathbf Q(t+T))-V(\mathbf Q(t))\mid\mathcal F_t\right]
\le
BT^2-\delta_T\|\mathbf Q(t)\|_1,
\qquad
\delta_T=\chi\Delta_T(\rho)>0.
\]
Summing this inequality over successive frame epochs and telescoping the Lyapunov terms gives a finite time-average expected queue at those epochs. Because one-slot queue increments are bounded, the queues within each finite frame differ from the frame-start queue by only a bounded amount. The same finiteness therefore extends to the all-slot time-average expected total movement queue stated in the corollary.
\end{proof}

When $I=0$, the condition in Corollary~\ref{cor:fixedT_stability} reduces to the classical admissibility requirement $\rho<1$; for $I>0$, the same stability mechanism is retained after accounting explicitly for the service capacity consumed by signal transitions.

Corollary~\ref{cor:fixedT_stability} makes explicit that $T>T_c(\rho)$ is a sufficient strong-stability condition for the fixed-timescale controller. Lemma~\ref{lem:Tslot_drift} also makes the two effects of $T$ transparent. The first term, $BT^2$, collects the usual bounded quadratic remainder together with the cumulative error caused by holding a phase chosen from the frame-initial queue state. At slot $t+\tau$, the traffic state can differ from the state used for the decision by at most $O(\tau)$; summing this discrepancy across the frame gives $O(T^2)$. Thus, a longer frame gradually makes the active phase less representative of the traffic state encountered later in that frame.

The second term describes the opposite effect. Without switching loss, a normalized load $\rho$ leaves a nominal reserve proportional to $(1-\rho)T$ over a frame of duration $T$. Reserving $I$ units of effective service time for a possible transition leaves the switching-adjusted frame slack
\begin{equation}
\Delta_T(\rho)
=
(1-\rho)T-I.
\label{eq:frame_service_slack}
\end{equation}
By Proposition~\ref{prop:critical_timescale}, this quantity is positive exactly when $T>T_c(\rho)$. Increasing $T$ therefore creates more usable service margin, but it does so only linearly, whereas the within-frame state-evolution term grows quadratically. Equation~\eqref{eq:Tslot_drift} is therefore the point at which the responsiveness--switching-efficiency trade-off becomes explicit in the Lyapunov analysis.

\begin{proposition}[Fixed-timescale queue bound with switching loss]
\label{prop:queue_bound}
Under the conditions of Lemma~\ref{lem:Tslot_drift}, the long-run all-slot average total queue satisfies
\begin{equation}
\overline{Q}(T)
\le
K_Q
\frac{T^2}
{(1-\rho)T-I},
\label{eq:queue_bound}
\end{equation}
for some constant $K_Q>0$ independent of $T$.
\end{proposition}

\begin{proof}
Let $t_k=kT$. Summing Eq.~\eqref{eq:Tslot_drift} over $K$ consecutive frames gives
\begin{equation}
\chi\Delta_T(\rho)
\sum_{k=0}^{K-1}
\mathbb{E}\!\left[\|\mathbf{Q}(t_k)\|_1\right]
\le
KBT^2
+
\mathbb{E}[V(\mathbf{Q}(0))].
\label{eq:frame_epoch_bound}
\end{equation}
After division by $K\chi\Delta_T(\rho)$ and passage to the limit, the average queue observed at frame boundaries is at most $(B/\chi)T^2/\Delta_T(\rho)$. This is not yet the all-slot average because queues continue to change within each frame. Bounded queue increments give
\[
\|\mathbf Q(t_k+\tau)\|_1
\le
\|\mathbf Q(t_k)\|_1+G\tau,
\qquad 0\le\tau<T.
\]
Averaging this inequality over the slots of every frame adds at most $G(T-1)/2$. Since $0<\Delta_T(\rho)<T$, we have $T\le T^2/\Delta_T(\rho)$, so this within-frame correction can be absorbed into the same scaling. This yields Eq.~\eqref{eq:queue_bound} with a $T$-independent constant $K_Q$.
\end{proof}

Equation~\eqref{eq:queue_bound} turns the two terms in Lemma~\ref{lem:Tslot_drift} into a single performance envelope. Its denominator is the service slack available to dissipate queues, while its numerator reflects the frame-length penalty incurred by retaining a decision as the traffic state evolves. As $T$ approaches $T_c(\rho)$ from above, the denominator approaches zero and the bound becomes large even though the system has just entered the conservative stable region. Moving away from that boundary initially improves the bound because service slack is restored faster than the frame penalty grows. For sufficiently long frames, however, the quadratic numerator dominates.

To locate the turning point, define $F(T)=T^2/[(1-\rho)T-I]$ for $T>T_c(\rho)$. Its derivative is
\[
F'(T)
=
\frac{T[(1-\rho)T-2I]}{[(1-\rho)T-I]^2}.
\]
The sign changes only when $(1-\rho)T=2I$. Hence the unique minimizer of this analytical bound is
\begin{equation}
T_U(\rho)
=
\frac{2I}{1-\rho}
=
2T_c(\rho).
\label{eq:upper_timescale}
\end{equation}
For $T_c<T<T_U$, increasing the frame length improves the analytical bound because the gain in switching-adjusted service slack dominates the additional state-evolution penalty. Once $T>T_U$, the quadratic frame penalty dominates and the bound increases. The two characteristic timescales therefore motivate the performance-oriented interval
\begin{equation}
T_c(\rho)<T\le2T_c(\rho).
\label{eq:operating_range}
\end{equation}
Here, $T_c$ defines the stability-oriented lower timescale, while $2T_c$ provides the performance-oriented characteristic timescale that minimizes the derived queue-bound envelope. Together they identify a load-dependent operating range for balancing switching efficiency and state responsiveness. The derivative treats $T$ as a continuous analytical variable, while an implementation with discrete decision intervals can select the nearest admissible value.

\begin{figure}
    \centering
    \includegraphics[width=0.6\linewidth]{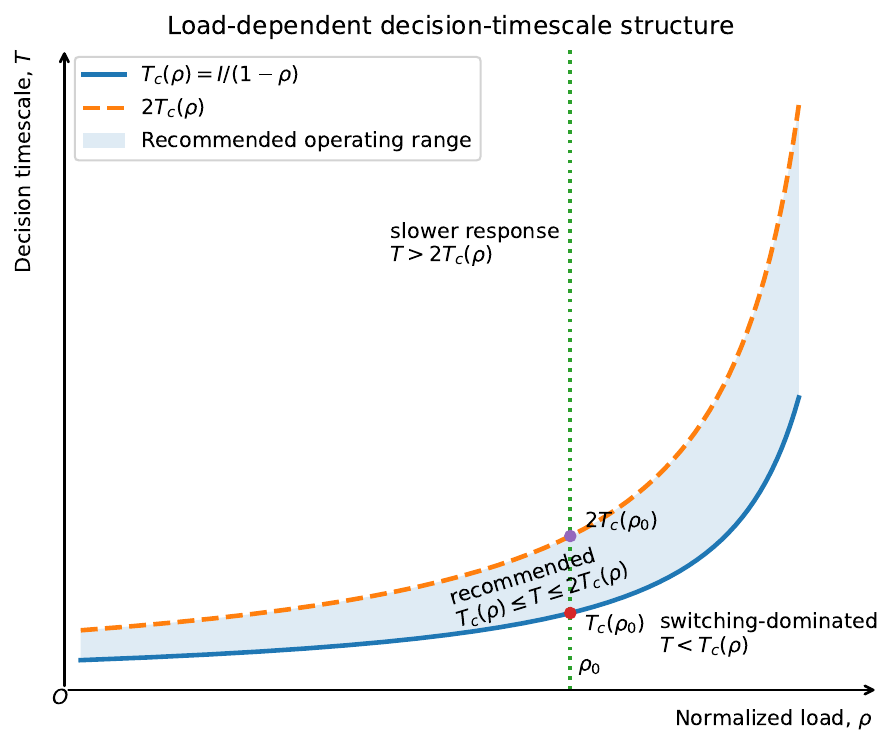}
    \caption{Load-dependent decision-timescale structure. The critical timescale $T_c(\rho)=I/(1-\rho)$ defines the lower boundary for a positive switching-adjusted service reserve, while $2T_c(\rho)$ gives the performance-oriented characteristic scale from the queue-bound analysis, where $\rho<1$. The shaded band indicates $T_c(\rho)<T\le2T_c(\rho)$, illustrated at an arbitrary operating load $\rho_0$.}
    \label{fig:timescale_structure}
\end{figure}

Figure~\ref{fig:timescale_structure} summarizes the resulting timescale structure in the $(\rho,T)$ plane. The lower curve is the critical timescale $T_c(\rho)$, below which the switching-adjusted service reserve is nonpositive, while the upper curve $2T_c(\rho)$ marks the characteristic minimizer of the analytical queue-bound envelope. For an arbitrary operating load $\rho_0$, the corresponding interval $T_c(\rho_0)<T\le2T_c(\rho_0)$ is the performance-oriented range. Both characteristic timescales increase with $\rho$, showing that the control interval should lengthen as the available service reserve becomes smaller; the switching loss $I$ scales both curves proportionally.

\subsection{Normalized decision-timescale policy}
\label{subsec:normalized_policy}

The preceding results suggest expressing the decision interval relative to the critical timescale:
\begin{equation}
T(\rho)
=
\kappa T_c(\rho)
=
\kappa\frac{I}{1-\rho},
\qquad
1<\kappa\le2.
\label{eq:normalized_policy}
\end{equation}
The dimensionless factor $\kappa=T/T_c$ specifies where the controller operates relative to the critical boundary rather than prescribing the same physical interval under every traffic condition. Substituting Eq.~\eqref{eq:normalized_policy} into the frame slack gives $\Delta_T(\rho)=(\kappa-1)I>0$. Thus, every $\kappa>1$ preserves a positive conservative service margin, while $1<\kappa\le2$ places the operating point between the critical boundary and the characteristic minimizer of the queue-bound envelope. The main theoretical implication is the scaling $T(\rho)\propto I/(1-\rho)$: decision intervals should increase when switching becomes more costly or when the available service reserve becomes smaller.

\section{Variable-Timescale Max-Pressure Control}
\label{sec:vtmp}

Section~\ref{sec:theory} derives how the appropriate MP decision timescale changes with traffic load when one interval is shared by the network. The next step is to remove that common-timescale restriction. VT-MP does so without changing the pressure definition or the phase-ranking rule: it localizes the load measure, applies the same normalized timing principle at each intersection, and allows the resulting update times to be nonaligned. This section first presents that local control rule and its physical interpretation, then shows how the classical MP stability mechanism extends to asynchronous heterogeneous updates, and finally introduces a finite-storage correction for microscopic traffic networks.

\subsection{Local formulation of VT-MP}
\label{subsec:local_vtmp}

The common-timescale analysis in Section~\ref{sec:theory} links the decision interval to a network-level load $\rho$, but urban intersections rarely operate at the same degree of saturation at the same time. VT-MP therefore applies the same normalized timing principle locally. The construction requires two quantities at each intersection: a local service region describing what the signal can provide through time sharing among its feasible phases, and an offered movement-flow vector describing the traffic that currently needs that service. Their ratio defines the local load. The controller then maps that load to a local decision interval. Importantly, the pressure definition and phase-ranking rule remain those of conventional MP; VT-MP adds only the temporal decision of how long the selected phase should be retained before pressure is reconsidered.

For intersection $n$, define the local nominal service region as the downward closure of the convex hull of its feasible phase-service vectors,
\begin{equation}
\mathcal{S}_n
=
\left\{
\mathbf{s}_n\ge0:
\mathbf{s}_n
\le
\sum_{p\in\mathcal{P}_n}\theta_p\boldsymbol{\mu}_n(p),
\;
\theta_p\ge0,
\;
\sum_{p\in\mathcal{P}_n}\theta_p=1
\right\},
\label{eq:local_service_region}
\end{equation}
where $\boldsymbol{\mu}_n(p)$ collects the saturation service rates of the movements in $\mathcal M_n$. The coefficients $\theta_p$ represent fractions of productive time assigned to the feasible phases. The downward closure simply recognizes that a feasible signal plan can always use less than the maximum service available on one or more movements.

Let $\boldsymbol\nu_{n,k}$ denote the movement-rate vector offered to intersection $n$ during its $k$th local decision frame. This flow includes both external demand and traffic routed from upstream intersections. The local normalized load is defined by
\begin{equation}
\rho_{n,k}
=
\inf\left\{
r>0:
\boldsymbol\nu_{n,k}\in r\mathcal S_n
\right\}.
\label{eq:local_load}
\end{equation}
Thus, $\rho_{n,k}$ has the same interpretation locally that $\rho$ has at the network level: it is the fraction of the local service region required to support the offered movement flow. A small value indicates substantial local reserve, whereas a value near one indicates that the intersection is operating close to its nominal service boundary. Because $\rho_{n,k}$ can vary with both $n$ and $k$, it provides the mechanism through which VT-MP responds to spatially and temporally heterogeneous traffic conditions.

The core VT-MP timing rule follows directly by applying the normalized policy in Eq.~\eqref{eq:normalized_policy} to the local load. If $I_n>0$ is the equivalent switching loss at intersection $n$, then
\begin{equation}
\boxed{
T_{n,k}
=
\kappa\frac{I_n}{1-\rho_{n,k}},
\qquad
1<\kappa\le2.}
\label{eq:local_timescale}
\end{equation}
Equation~\eqref{eq:local_timescale}, together with the local-load definition in Eq.~\eqref{eq:local_load}, is the central mechanism of VT-MP. At update epoch $t_{n,k}$, intersection $n$ computes its current movement pressures and chooses the conventional MP phase,
\[\boxed{
p_{n,k}
=
\arg\max_{p\in\mathcal P_n}
\sum_{(a,b)\in\mathcal M_n}
\mu_{ab}(p)w_{ab}(t_{n,k}),
\qquad
t_{n,k+1}=t_{n,k}+T_{n,k}.}
\]
The selected phase is retained over the local frame, apart from the transition interval $I_n$ when a phase change occurs. The method therefore augments the conventional MP decision ``which phase?'' with a second local decision ``for how long?'' Because the interval is determined independently at each node, local update epochs generally do not coincide and the network operates asynchronously without imposing a common control clock.

The timing rule also gives a useful normalization property. For frame $(n,k)$, define the switching-adjusted local slack as $\Delta_{n,k}=(1-\rho_{n,k})T_{n,k}-I_n$, and let $T_{c,n,k}=I_n/(1-\rho_{n,k})$ denote the corresponding local critical timescale.

\begin{proposition}[Uniform local service margin under VT-MP]
\label{prop:local_margin}
Under Eq.~\eqref{eq:local_timescale},
\begin{equation}
\Delta_{n,k}
=
(\kappa-1)I_n,
\qquad
\frac{T_{n,k}}{T_{c,n,k}}
=
\kappa.
\label{eq:local_margin_result}
\end{equation}
The fractional reserve is $\sigma_{n,k}=\Delta_{n,k}/T_{n,k}=((\kappa-1)/\kappa)(1-\rho_{n,k})$. Hence, if $I_n\ge I_{\min}>0$ and $\rho_{n,k}\le\bar\rho<1$, both the absolute and fractional service margins are uniformly positive over the network and time.
\end{proposition}

\begin{proof}
Substituting Eq.~\eqref{eq:local_timescale} into the slack definition gives
\[
\Delta_{n,k}
=(1-\rho_{n,k})
\kappa\frac{I_n}{1-\rho_{n,k}}-I_n
=(\kappa-1)I_n.
\]
The ratio $T_{n,k}/T_{c,n,k}=\kappa$ follows from the same substitution, and division by $T_{n,k}$ gives the stated fractional reserve. The lower bounds then follow directly from $I_n\ge I_{\min}$ and $\rho_{n,k}\le\bar\rho$.
\end{proof}

Proposition~\ref{prop:local_margin} gives the main interpretation of the local rule. VT-MP does not attempt to equalize the number of seconds used at different intersections. Instead, every node operates at the same normalized position $\kappa$ relative to its own critical timescale. A heavily loaded intersection therefore uses a longer physical interval than a lightly loaded one, while both retain a comparable normalized margin above their local stability boundary. This normalization has two important consequences. First, it makes the timing rule directly comparable across intersections with different loads and switching losses, because the control interval is expressed relative to the locally relevant stability scale rather than in absolute seconds. Second, it allows spatial heterogeneity to emerge automatically from local operating conditions: heavily loaded intersections reduce the relative impact of switching losses through longer frames, whereas lightly loaded intersections retain more frequent state updates. At the same time, the positive normalized service margin is preserved at every node, providing a common analytical structure for the asynchronous heterogeneous stability result developed next.

\subsection{Stability and performance implications}
\label{subsec:vtmp_theory}

The local rule creates different decision intervals and nonaligned update epochs across the network. We first establish that this asynchronous operation preserves the classical MP stability mechanism and then examine the performance implications of removing the common-timescale constraint. The definition of strong stability remains exactly the one introduced in Section~\ref{subsec:classical_mp_stability}; what changes is the time horizon over which the Lyapunov drift must be evaluated. Because intersections no longer share a common one-slot update epoch, the stability analysis first needs a finite-horizon counterpart of the classical one-step drift argument and then must show that asynchronous VT-MP satisfies it.

\subsubsection{Stability under asynchronous heterogeneous updates}

The following lemma provides the first step. It translates a negative drift over any fixed finite analysis horizon into the same strong-stability notion used for conventional MP. The horizon is an analytical device rather than a common signal-control interval.

\begin{lemma}[Finite-horizon stability criterion]
\label{lem:drift_strong_stability}
Suppose queue increments are uniformly bounded. If there exist a finite integer $H\ge1$ and constants $B_H<\infty$ and $\delta_H>0$ such that
\begin{equation}
\mathbb E\!\left[
V(\mathbf Q(t+H))-V(\mathbf Q(t))
\mid\mathcal F_t
\right]
\le
B_H-\delta_H\|\mathbf Q(t)\|_1
\label{eq:general_H_drift}
\end{equation}
for all $t$, then the queue process is strongly stable in the sense of Definition~\ref{def:strong_stability}.
\end{lemma}

\begin{proof}
Apply Eq.~\eqref{eq:general_H_drift} at the sampled epochs $0,H,2H,\ldots,(K-1)H$ and sum the resulting inequalities. The Lyapunov terms telescope, giving
\[
\delta_H\sum_{k=0}^{K-1}
\mathbb E\!\left[\|\mathbf Q(kH)\|_1\right]
\le
KB_H+\mathbb E[V(\mathbf Q(0))].
\]
After division by $K\delta_H$ and passage to the limit, the expected total queue is finite on average at the sampled epochs. Because one-slot queue increments are uniformly bounded and $H$ is finite, the queues between two sampled epochs can differ from the preceding sampled state by only a finite amount. The all-slot average in Eq.~\eqref{eq:strong_stability_definition} is therefore finite as well.
\end{proof}

Lemma~\ref{lem:drift_strong_stability} does not impose a common signal-control interval. The remaining task is therefore specific to VT-MP: to show that local frames with different lengths still produce a positive queue-dependent service reserve over a common analysis horizon, while the loss caused by nonaligned updates remains bounded.

To isolate this asynchronous effect, first consider a fixed mean arrival vector $\boldsymbol\lambda$ and the routing-consistent effective movement flow $\boldsymbol\nu$ from Eq.~\eqref{eq:effective_flow_balance}. Let $\boldsymbol\nu_n$ be its restriction to $\mathcal M_n$, define $\rho_n$ through Eq.~\eqref{eq:local_load}, and assume a uniform local reserve $0\le\rho_n\le\bar\rho<1$ for every node. This is the local counterpart of strict admissibility in Definition~\ref{def:capacity_region}. Under this operating condition, node $n$ uses
\begin{equation}
T_n
=
\kappa\frac{I_n}{1-\rho_n}.
\label{eq:stationary_local_timescale}
\end{equation}
The intersections may start their local frames at arbitrary, nonaligned epochs. Assume also $0<I_{\min}\le I_n\le I_{\max}<\infty$. The local intervals are then uniformly bounded,
\[
0<T_{\min}\le T_n\le T_{\max}<\infty,
\qquad
T_{\min}=\kappa I_{\min},
\quad
T_{\max}=\frac{\kappa I_{\max}}{1-\bar\rho}.
\]

For the network drift, the external arrival term must be expressed in the same pressure coordinates used by the MP decision. The classical flow--pressure identity in Eq.~\eqref{eq:flow_pressure_identity} performs exactly this conversion: internally routed flows cancel, allowing the node-wise pressure--service comparisons below to combine into a network-level Lyapunov argument.

\begin{theorem}[Strong stability of asynchronous heterogeneous VT-MP]
\label{thm:async_stability}
Under the bounded-arrival and bounded-service assumptions of Section~\ref{subsec:network_model}, open routing, the uniformly admissible local loads above, and the VT-MP intervals in Eq.~\eqref{eq:stationary_local_timescale}, the asynchronous heterogeneous VT-MP policy is strongly stable. In particular, there exist a finite analysis horizon $H$ and constants $B_H<\infty$ and $\delta_H>0$ satisfying Eq.~\eqref{eq:general_H_drift}.
\end{theorem}

\begin{proof}
The proof has three steps. First, the local timing rule guarantees a positive service reserve at every intersection. Let $\phi_n=1-I_n/T_n$. From Eq.~\eqref{eq:stationary_local_timescale},
\[
\phi_n-\rho_n
=
\frac{\kappa-1}{\kappa}(1-\rho_n)
\ge
\sigma_{\min},
\qquad
\sigma_{\min}
=
\frac{\kappa-1}{\kappa}(1-\bar\rho)>0.
\]
Using the convexity and downward closure of $\mathcal S_n$, this reserve yields a pressure-compatible comparison service $\bar{\mathbf s}_n^{\,\mathbf w}\in\phi_n\mathcal S_n$ such that, for some $d_{\min}>0$ independent of the queue state,
\begin{equation}
\mathbf w_n^{\top}\boldsymbol\nu_n
-
\mathbf w_n^{\top}\bar{\mathbf s}_n^{\,\mathbf w}
\le
-\sigma_{\min}d_{\min}\|\mathbf w_n^{+}\|_1.
\label{eq:local_pressure_reserve}
\end{equation}
Summing Eq.~\eqref{eq:local_pressure_reserve} over intersections, using Eq.~\eqref{eq:flow_pressure_identity}, and applying the open-routing pressure-to-queue bound in Eq.~\eqref{eq:pressure_queue_norm} converts this local service reserve into a network stabilizing term proportional to $-\|\mathbf Q\|_1$.

Second, asynchrony causes each retained phase to be based on a traffic state of bounded age. If $\tau_n(s)$ is the update epoch at which the phase executed by node $n$ at slot $s$ was selected, then $0\le s-\tau_n(s)\le T_{\max}$. Bounded queue increments and the linearity of pressure imply, for some finite $L_P$,
\begin{equation}
P_{n,p_n(s)}(s)
\ge
\max_{p\in\mathcal P_n}P_{n,p}(s)
-
2L_PT_{\max}
\label{eq:approximate_mp}
\end{equation}
whenever node $n$ is in a productive slot. Thus the loss caused by using a phase selected from an earlier local state is bounded independently of the queue magnitude, whereas the service-reserve term in Eq.~\eqref{eq:local_pressure_reserve} grows with the queues.

Third, evaluate the network over a common analysis horizon $H>2T_{\max}$. At each node, all but the local frames intersecting the two ends of the horizon are complete. The complete frames contribute the queue-dependent negative term from Eq.~\eqref{eq:local_pressure_reserve}; the two short end portions, the stale-pressure error in Eq.~\eqref{eq:approximate_mp}, and the usual quadratic and within-horizon state-change terms are bounded. Consequently, for finite constants $c_1,c_2>0$ and $\widetilde B_H<\infty$,
\[
\mathbb E
\left[
V(\mathbf Q(t+H))-V(\mathbf Q(t))
\mid\mathcal F_t
\right]
\le
\widetilde B_H
-
\big[c_1(H-2T_{\max})-c_2T_{\max}\big]
\|\mathbf Q(t)\|_1.
\]
Because $c_1>0$, a finite $H$ can be selected so that the coefficient in brackets is positive. The resulting inequality has the form of Eq.~\eqref{eq:general_H_drift}; Lemma~\ref{lem:drift_strong_stability} then gives strong stability. Appendix~\ref{app:async_stability} supplies the complete-frame decomposition and explicit bounds absorbed into $c_1$, $c_2$, and $\widetilde B_H$.
\end{proof}

Theorem~\ref{thm:async_stability} shows why asynchronous timing does not destroy the MP stability mechanism. The local timing rule preserves a positive service reserve, while bounded decision intervals limit how old the pressure information used by any intersection can become. The former produces a stabilizing term that grows with the queue magnitude; the latter contributes only bounded errors. Once the network is viewed over a sufficiently long common analysis horizon, the stabilizing effect dominates. The horizon $H$ is used only for analysis and is not a common signal-control interval imposed on the intersections.

The preceding strong-stability result, established under a fixed mean arrival vector, can be extended to piecewise-stationary traffic demand whose mean changes between operating periods but remains approximately constant within each period.

\begin{corollary}[Strong stability under piecewise-stationary traffic demand]
\label{cor:piecewise_stationary}
Let $0=\tau_0<\tau_1<\tau_2<\cdots$ define traffic operating periods $\mathcal I_m=[\tau_m,\tau_{m+1})$. Within period $m$, suppose the exogenous arrivals have fixed conditional mean $\boldsymbol\lambda^{(m)}$ and the induced local loads satisfy $\rho_n^{(m)}\le\bar\rho<1$ for all intersections. After its first update in that period, node $n$ uses $T_n^{(m)}=\kappa I_n/(1-\rho_n^{(m)})$. Then the constants in Theorem~\ref{thm:async_stability} can be chosen uniformly across operating periods. Any $H$-window that begins after the local adjustment time and ends before the next change in mean demand satisfies Eq.~\eqref{eq:general_H_drift} with the same positive $\delta_H$. Moreover, if every operating period lasts longer than a finite threshold $L^\star$, the piecewise-stationary system is strongly stable.
\end{corollary}

\begin{proof}
The uniform load bound gives the same finite $T_{\max}$ and the same positive service reserve in every operating period. Following a change in mean arrival rate, an intersection may still be completing one interval chosen under the previous traffic condition, but all intersections update to the new condition within at most $T_{\max}$ slots. Every complete $H$-window after this adjustment therefore satisfies
\[
\mathbb E
\left[
V(\mathbf Q(s_j+H))-V(\mathbf Q(s_j))
\mid\mathcal F_{s_j}
\right]
\le
B_H-\delta_H\|\mathbf Q(s_j)\|_1.
\]
If an operating period contains $J$ such complete windows, summing these inequalities gives
\[
\sum_{j=0}^{J-1}
\mathbb E[\Delta_HV(s_j)\mid\mathcal F_{s_j}]
\le
JB_H
-
\delta_H
\sum_{j=0}^{J-1}\|\mathbf Q(s_j)\|_1.
\]
The initial adjustment interval and any remaining interval shorter than $H$ at the end of the period have uniformly bounded duration, so their additional drift is bounded by a finite boundary term. For sufficiently large $J$, the queue-dependent negative term accumulated over the complete windows dominates this bounded effect. Hence there exists a finite minimum operating-period duration $L^\star$ for which the drift over successive operating periods remains negative outside a bounded set, and Lemma~\ref{lem:drift_strong_stability} gives strong stability. The explicit boundary bounds and the construction of $L^\star$ are provided in Appendix~\ref{app:async_stability}.
\end{proof}

The corollary has a simple traffic interpretation. A change in mean demand does not require all intersections to update at the same instant. Because every local interval is bounded, the network completes its adjustment within a finite time. During the stationary part of each sufficiently long operating period, the same negative-drift mechanism as in Theorem~\ref{thm:async_stability} applies repeatedly, and the negative-drift contribution accumulated during these stationary portions dominates the bounded effect of the short adjustment intervals.

\subsubsection{Performance implication of heterogeneous timescales}

The stability results above establish that local, nonaligned decision intervals preserve the network-level MP stability mechanism. We next examine the performance implication of allowing these intervals to differ when traffic loads vary spatially.
Under a common decision interval $T^{\mathrm c}$, every intersection retains its selected phase until the next network-wide decision epoch. To maintain a positive conservative switching-adjusted service margin at every node, the common interval must satisfy
\begin{equation}
T^{\mathrm c}
>
\max_{n\in\mathcal V}
\frac{I_n}{1-\rho_n}.
\label{eq:common_timescale_requirement}
\end{equation}
Hence, the common timescale is determined by the intersection with the largest local critical timescale. When local loads are heterogeneous, this interval can be substantially longer than the local critical timescale at a lightly loaded intersection.

Such a long holding interval can reduce service utilization. If the movements served by the selected phase become empty before the next common decision epoch while another phase has queued demand, the remaining potential service cannot be reallocated until the common interval expires. A shorter local interval can instead trigger an earlier MP update and use the remaining time to serve another phase.

\begin{proposition}[Strict service advantage of heterogeneous timing]
\label{prop:heterogeneous_service_dominance}
Consider an intersection over a common comparison horizon $T^{\mathrm c}$, while VT-MP uses a shorter local interval $T_n<T^{\mathrm c}$. Suppose that the movements served by the initially selected phase remain empty over $[T_n,T^{\mathrm c})$, and VT-MP selects at $T_n$ another feasible phase with sufficient queued demand. If
\begin{equation}
T^{\mathrm c}-T_n>I_n,
\label{eq:strict_service_condition}
\end{equation}
then
\begin{equation}
D_n^{\mathrm{het}}(T^{\mathrm c})
>
D_n^{\mathrm{com}}(T^{\mathrm c}).
\label{eq:strict_service_dominance}
\end{equation}
\end{proposition}

\begin{proof}
The two controllers generate the same realized service before $T_n$. Over $[T_n,T^{\mathrm c})$, the common-timescale controller retains a phase whose served movements remain empty and therefore generates no additional realized departure from that phase. VT-MP instead switches to a queued phase and, after the switching loss $I_n$, obtains positive realized service over the remaining duration $T^{\mathrm c}-T_n-I_n>0$. Hence its cumulative realized departure over the comparison horizon is strictly larger.
\end{proof}

At the network level, if at least one intersection satisfies Proposition~\ref{prop:heterogeneous_service_dominance} while the cumulative departures at the remaining intersections are no smaller under heterogeneous timing over the same comparison horizon, then
\[
\sum_{n\in\mathcal V}
D_n^{\mathrm{het}}
>
\sum_{n\in\mathcal V}
D_n^{\mathrm{com}}.
\]
Thus, an otherwise unused local service opportunity can translate directly into a strict increase in network-wide cumulative realized departures.

Proposition~\ref{prop:heterogeneous_service_dominance} identifies the service-utilization benefit of spatially heterogeneous timescales. A long interval can be desirable at a heavily loaded intersection because it reduces the relative effect of switching loss, whereas the same interval may cause a lightly loaded intersection to retain a phase after its demand has been served. Heterogeneous timing allows heavily loaded nodes to retain longer frames while lightly loaded nodes reconsider their phase earlier and reallocate otherwise unused service opportunities.

The same spatial mismatch also appears in the analytical queue-bound structure. For a fixed operating condition, suppress the local update index $k$ and write $\rho_n$ and $T_n$ for the corresponding node-level quantities. Proposition~\ref{prop:queue_bound} suggests the local surrogate
\begin{equation}
F_n(T_n)
=
K_n
\frac{T_n^2}
{(1-\rho_n)T_n-I_n},
\qquad
T_n>\frac{I_n}{1-\rho_n},
\label{eq:local_bound_surrogate}
\end{equation}
where $K_n>0$ collects node-dependent bounded constants. Define
\[
\mathcal T_{\mathrm{het}}
=
\prod_{n\in\mathcal V}
\left(
\frac{I_n}{1-\rho_n},
\infty
\right),
\qquad
\mathcal T_{\mathrm{com}}
=
\left\{
\mathbf T\in\mathcal T_{\mathrm{het}}:
T_n=T_m,\ \forall n,m
\right\}.
\]
The first permits each intersection to choose its own admissible interval, whereas the second imposes one common interval network-wide.

\begin{proposition}[Analytical performance dominance of heterogeneous timescales]
\label{prop:heterogeneous_bound}
For the separable surrogate in Eq.~\eqref{eq:local_bound_surrogate},
\begin{equation}
\min_{\mathbf T\in\mathcal T_{\mathrm{het}}}
\sum_{n\in\mathcal V}F_n(T_n)
\le
\min_{\mathbf T\in\mathcal T_{\mathrm{com}}}
\sum_{n\in\mathcal V}F_n(T_n).
\label{eq:heterogeneous_bound}
\end{equation}
\end{proposition}

\begin{proof}
Since $\mathcal T_{\mathrm{com}}\subseteq\mathcal T_{\mathrm{het}}$, minimizing the same objective over the larger heterogeneous feasible set cannot yield a larger optimum.
\end{proof}

The two propositions capture complementary benefits of heterogeneous timing. Proposition~\ref{prop:heterogeneous_service_dominance} establishes a strict realized-service gain when a common interval retains a locally depleted phase beyond the point at which an alternative queued phase could be served. Proposition~\ref{prop:heterogeneous_bound} shows more generally that removing the common-timescale constraint cannot increase the minimum analytical queue-bound surrogate. When local loads are similar, their characteristic timescales are also similar and the common-timescale restriction is mild. As spatial load heterogeneity increases, these characteristic timescales separate: a common interval is increasingly governed by the most heavily loaded nodes, whereas VT-MP allows each intersection to operate at its own load-dependent timescale.

\subsection{Finite-storage implementation}
\label{subsec:finite_storage}

The preceding analysis follows the standard infinite-storage store-and-forward model, under which an upstream movement can discharge whenever it receives service. A microscopic road network introduces an additional physical limitation: a downstream link may become full before the planned VT-MP interval ends, an effect that has motivated capacity-aware and downstream-capacity extensions of backpressure signal control \citep{gregoire2015capacity,hao2020downstream, li2019position, tan2026cvmp}. In that case, retaining the planned phase for its full theoretical duration could prescribe a discharge that exceeds the available downstream receiving capacity. We therefore treat the theoretical interval as a planned timescale and apply a local receiving-supply check to determine how much of that interval can actually be executed.

%\subsubsection{Downstream receiving-supply constraint}

Let $K_j$ denote the storage capacity of downstream link $j$ and $N_j(t)$ its current occupancy. With a usable-storage factor $\eta\in(0,1]$, define the available receiving supply as
\begin{equation}
S_j^{\mathrm{rec}}(t)
=
\left[
\eta K_j-N_j(t)
\right]_+.
\label{eq:receiving_supply}
\end{equation}
Let $\widehat D_m(T)$ denote the predicted cumulative discharge from upstream movement $m$ over a candidate interval of length $T$, and let $\beta_{mj}$ be the fraction of that discharge entering link $j$. The predicted cumulative inflow is
\[
Y_j(T)
=
\sum_m\beta_{mj}\widehat D_m(T).
\]
A candidate interval is receiving-feasible when $Y_j(T)\le S_j^{\mathrm{rec}}(t)$ for every affected downstream link. This condition does not change which phase has the highest MP pressure; it only prevents the planned holding time from extending beyond the downstream space available to receive the associated discharge.

%\subsubsection{Executable decision interval}

Let $T_{n,k}^{\mathrm{plan}}$ denote the VT-MP interval given by Eq.~\eqref{eq:local_timescale}. The interval executed in the finite-storage network is
\begin{equation}
T_{n,k}^{\mathrm{exec}}
=
\max
\left\{
T:
0<T\le T_{n,k}^{\mathrm{plan}},
\;
Y_j(T)\le S_j^{\mathrm{rec}}(t_{n,k})
\;\forall j
\right\}.
\label{eq:executable_interval}
\end{equation}
Thus $T_{n,k}^{\mathrm{plan}}$ is the interval suggested by the load-dependent VT-MP theory, whereas $T_{n,k}^{\mathrm{exec}}$ is the longest feasible duration not exceeding that interval that remains compatible with the currently available downstream storage. When sufficient receiving space is available, the two are identical; when downstream storage becomes restrictive, the executed interval is shortened. This receiving-supply correction belongs to the finite-storage implementation layer, while the analytical results above remain those of the infinite-storage store-and-forward model.

\section{Validation by Numerical and Simulation Experiments}
\label{sec:experiments}

The three experiments have complementary roles in validating the theory and evaluating the proposed controller. Experiment~1 is a mechanism experiment: it isolates the load--timescale relation at a single intersection and tests the theoretical prediction against a microscopic decision-interval sweep. The resulting comparison also provides the dimensionless timing factor $\kappa$ used in the subsequent experiments. Experiments~2 and~3 then evaluate network-level effectiveness. Experiment~2 uses a $5\times5$ store-and-forward network that closely matches the infinite-storage assumptions of the analysis, whereas Experiment~3 tests the complete VT-MP implementation in microscopic SUMO traffic with detector-based state estimation, finite link storage, and spillback. The underlying MP phase-selection rule is unchanged across the three experiments; the progression is from mechanism-level theory validation to network-level effectiveness under increasingly realistic traffic dynamics.

\begin{figure*}
\centering
\includegraphics[width=0.98\textwidth]{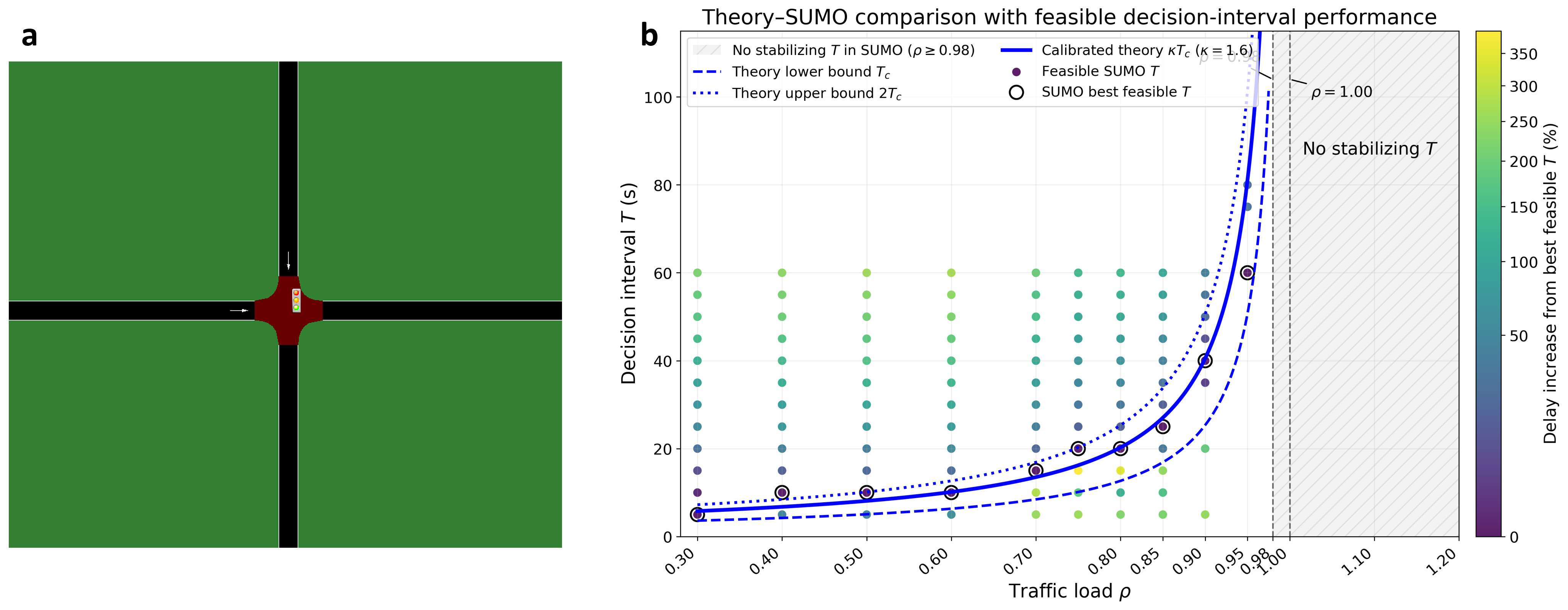}
\caption{Experiment~1: single-intersection mechanism and load-dependent decision timescale. (a) Microscopic SUMO scenario used to isolate the interaction between traffic load and the MP decision interval. (b) Comparison between the theoretical characteristic timescales and the microscopic decision-interval sweep. Colored points denote feasible tested intervals and their delay increase relative to the best feasible interval at the same load; black rings mark the best feasible interval. The blue curves show $T_c$, $2T_c$, and the calibrated policy $\kappa T_c$ with $\kappa=1.6$. The hatched region indicates the near-/over-capacity range in which no tested interval stabilizes the microscopic intersection.}
\label{fig:exp1_theory_sumo}
\end{figure*}

\subsection{Experiment 1: Single-intersection mechanism}
\label{subsec:exp_single}

The first experiment provides a mechanism-level validation of the decision-timescale theory by isolating the load--timescale relationship at a single intersection. Figure~\ref{fig:exp1_theory_sumo}(a) shows the microscopic SUMO scenario: an isolated four-arm intersection with two competing signal phases and symmetric horizontal and vertical demand. Deterministic uniform arrivals are used so that the normalized load can be varied without introducing directional imbalance. The tested range is $\rho=0.30$--$1.20$, with denser sampling near the nominal capacity boundary. Fixed decision intervals are tested primarily over $T=5$--$60$~s in 5-s increments, with additional longer intervals for selected near-capacity cases; each run lasts 3600~s.

The displayed yellow interval is 3~s. Continuous-green and forced-alternation pretrials are used to estimate the discharge capacity and the equivalent service loss associated with a phase change. These tests give an average saturation flow of approximately 2339~veh/h and an effective switching loss of $I_{\mathrm{eff}}=2.534$~s. The latter measures lost productive service rather than the physical yellow duration and is therefore the quantity used in the theoretical comparison.

Figure~\ref{fig:exp1_theory_sumo}(b) shows the resulting load--timescale relationship. The preferred feasible interval increases systematically with traffic load: short intervals of roughly 5--10~s are favored under light traffic, the preferred interval is about 20~s around $\rho=0.8$, and it reaches approximately 40~s by $\rho=0.90$. The persistent upward trend provides direct empirical validation of the theoretical mechanism. As the service reserve $1-\rho$ shrinks, repeated switching consumes a larger fraction of the available capacity, so retaining the selected phase for longer becomes increasingly beneficial, exactly as predicted by the scaling $T\propto I/(1-\rho)$.

Fitting the normalized policy over the subcritical cases up to $\rho=0.90$ gives $\kappa\approx1.6$, which is retained without further tuning in the subsequent network experiments. Over this range, the empirically preferred decision intervals lie broadly between the analytical scales $T_c$ and $2T_c$, supporting their interpretation as a stability-oriented lower boundary and a performance-oriented characteristic timescale. As the load approaches the nominal capacity boundary, however, the theoretical recommendation $I/(1-\rho)$ increases rapidly and the feasible microscopic timing range becomes increasingly restricted. At $\rho=0.95$, only three tested intervals remain feasible, although all still lie within the theoretical operating range, whereas at $\rho=0.98$ no tested interval prevents spillback. This near-capacity deviation from the point-queue prediction reflects the growing influence of finite storage and queue propagation in the microscopic model. Accordingly, $\rho=0.90$ is adopted in Experiment~3 as the operational upper range of the load-based recommendation, keeping the controller within the regime where the calibrated load--timescale relationship remains both numerically well behaved and operationally informative.

Overall, the single-intersection experiment validates the central theoretical prediction that the preferred MP decision timescale increases with traffic load and is governed by the derived characteristic scales. The experiment also supplies the normalized timing factor used in the network experiments and identifies the practically relevant load range for the microscopic implementation.

\subsection{Experiment 2: Store-and-forward network}
\label{subsec:exp_sf}

%\subsubsection{Network setting and demand pattern}

Having validated the load--timescale mechanism at a single intersection, the second experiment evaluates the network-level effectiveness of VT-MP when demand and the preferred decision timescale vary across both time and space. A $5\times5$ grid with 25 signalized intersections is used, with four feasible phases at each node. Traffic is represented by store-and-forward point queues with infinite storage and exact movement states, following the modeling tradition used in network-wide urban signal control \citep{diakaki2002tuc,aboudolas2009store}, so the experiment remains close to the assumptions used in the theoretical analysis. An explicit switching loss of $I=3$~s is imposed at every phase change.

\begin{figure*}
\centering
\includegraphics[width=0.96\textwidth]{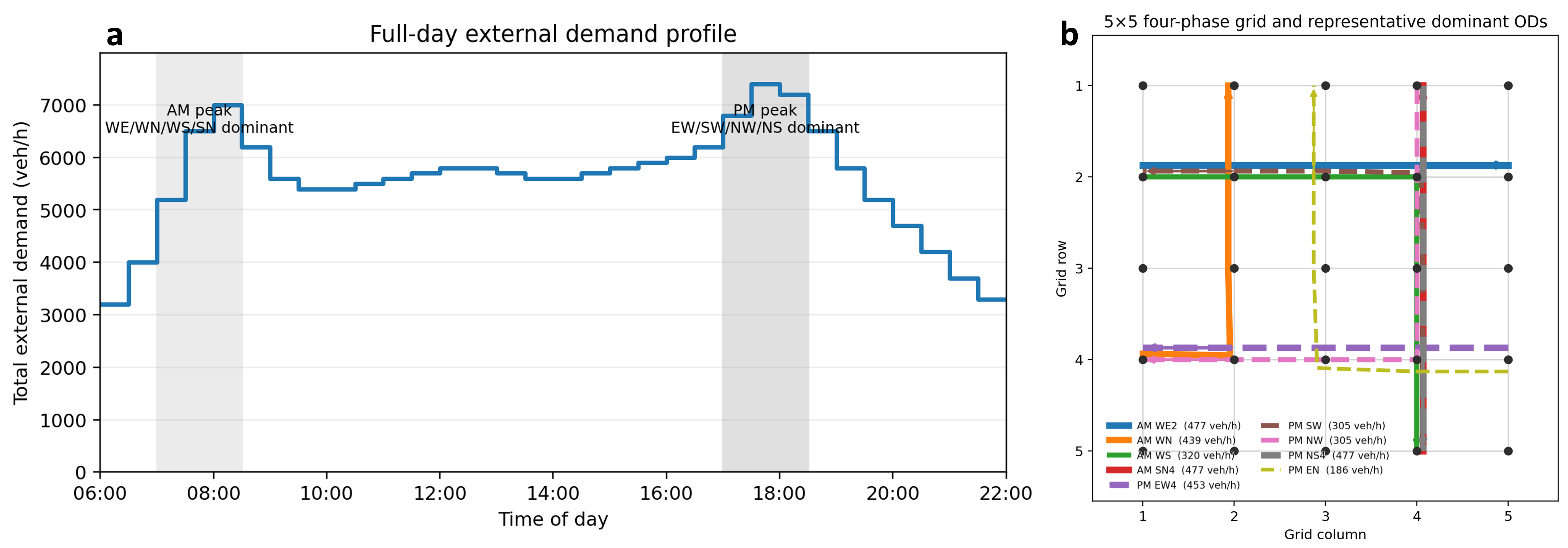}
\caption{Experiment~2 network and demand setting. Panel~(a) shows the full-day external demand profile, with the morning and evening peak periods highlighted. Panel~(b) shows the $5\times5$ four-phase grid and representative dominant OD paths during the two peak periods; line width is proportional to peak path demand, and arrows indicate travel direction.}
\label{fig:exp2_scenario}
\end{figure*}

Boundary traffic is distributed over 28 fixed paths. As shown in Fig.~\ref{fig:exp2_scenario}(a), the total external demand varies in 30-min blocks from 06:00 to 22:00, followed by a 30-min clearance period, and ranges from approximately 3200 to 7400~veh/h. The demand profile contains distinct morning and evening peaks separated by a relatively stable daytime period. The OD composition also changes over the day. Figure~\ref{fig:exp2_scenario}(b) shows the representative dominant OD paths during the two peak periods; line width is proportional to the corresponding peak path demand. The resulting combination of temporal demand variation and changing spatial flow concentration creates the spatiotemporally heterogeneous operating conditions required to test VT-MP.

The local normalized load is computed directly from the prescribed movement demand and known service rates. Decision intervals are quantized to a 5-s grid, and the value $\kappa=1.6$ calibrated in Experiment~1 is used without further tuning. Five timing strategies are compared:
\begin{itemize}
    \item \textit{Best fixed}: one network-wide interval selected from the full fixed-$T$ sensitivity analysis;
    \item \textit{Hourly-best fixed}: an offline temporal benchmark that selects the best common fixed interval separately for each hour;
    \item \textit{Common-max}: the maximum local VT-MP recommendation applied synchronously to all intersections;
    \item \textit{Common-Q85}: the 85th percentile of the local VT-MP recommendations applied synchronously to all intersections; and
    \item \textit{Heterogeneous VT-MP}: node-specific load-dependent intervals executed asynchronously.
\end{itemize}
The fixed-$T$ sensitivity covers 5--60~s in 5-s increments. Ten independent arrival realizations are used, with common random arrivals across methods within each realization. The main performance measure is total queueing time,
\[
H_Q=\int_0^{T_{\mathrm{sim}}}\sum_{(a,b)\in\mathcal M}Q_{ab}(t)\,\mathrm dt,
\]
supplemented by the mean and peak number of vehicles in the network and demand completion.

\subsubsection{Overall network performance}

\begin{figure*}
\centering
\includegraphics[width=0.98\textwidth]{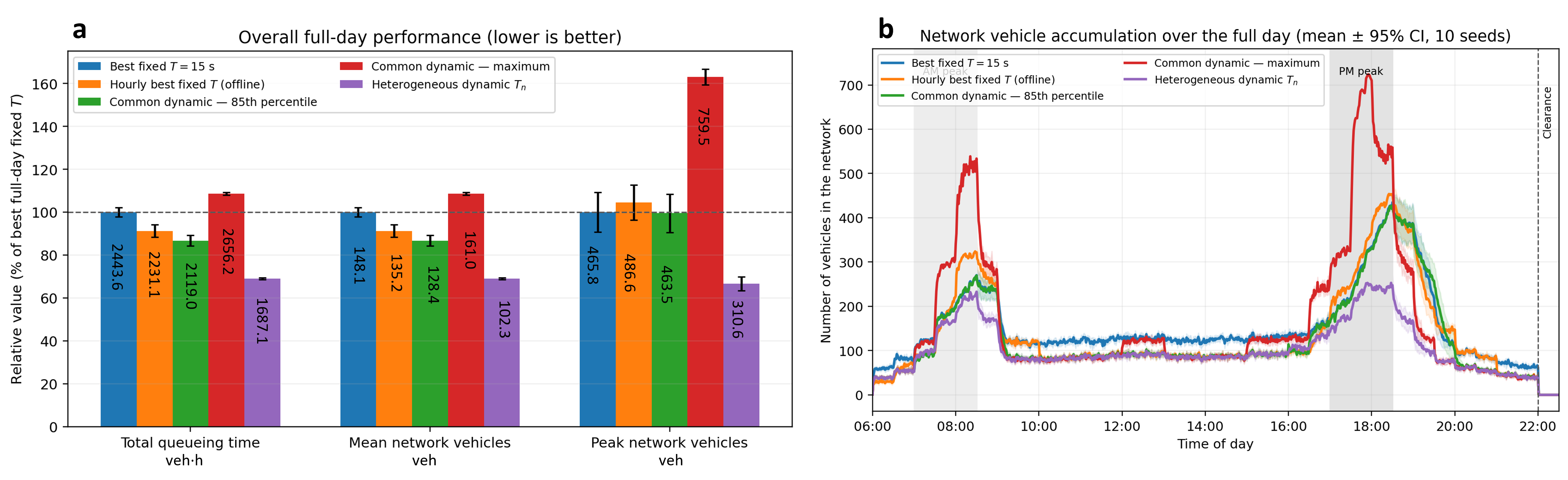}
\caption{Experiment~2 overall performance. Panel~(a) compares full-day total queueing time, mean network accumulation, and peak network accumulation relative to the best full-day fixed interval; the absolute values are shown inside the bars. Panel~(b) shows the evolution of network vehicle accumulation over the day, with lines denoting the mean across ten arrival realizations and shaded bands denoting the 95\% confidence interval. Lower values indicate better traffic performance.}
\label{fig:exp2_overall}
\end{figure*}

All five controllers serve the complete demand and clear the network, so their differences reflect congestion and queueing efficiency rather than differences in served demand. Figure~\ref{fig:exp2_overall} summarizes the full-day performance and the corresponding evolution of network vehicle accumulation. The heterogeneous VT-MP policy provides the strongest overall improvement: relative to the best fixed interval, total queueing time decreases from 2443.6 to 1687.1~veh$\cdot$h, a reduction of 31.0\%, while the mean network accumulation is reduced by the same proportion. The peak accumulation also falls substantially, from 465.8 to 310.6 vehicles.

The remaining common-timescale strategies show that temporal adaptation alone can be beneficial but does not guarantee improvement. Common-Q85 improves the full-day queueing performance relative to the best fixed interval, whereas Common-max performs worse and produces a pronounced accumulation peak during the evening demand surge. The hourly-best benchmark also improves the full-day queueing total but does not consistently reduce the peak accumulation. In contrast, heterogeneous VT-MP keeps the network accumulation lower through both peak periods and dissipates the accumulated traffic more rapidly afterward. The overall comparison therefore suggests that the main benefit does not arise simply from replacing one fixed interval with another time-varying network-wide interval; it comes from allowing the timing decision to follow local operating conditions.

\subsubsection{Temporal adaptation}

\begin{figure*}
\centering
\includegraphics[width=0.9\textwidth]{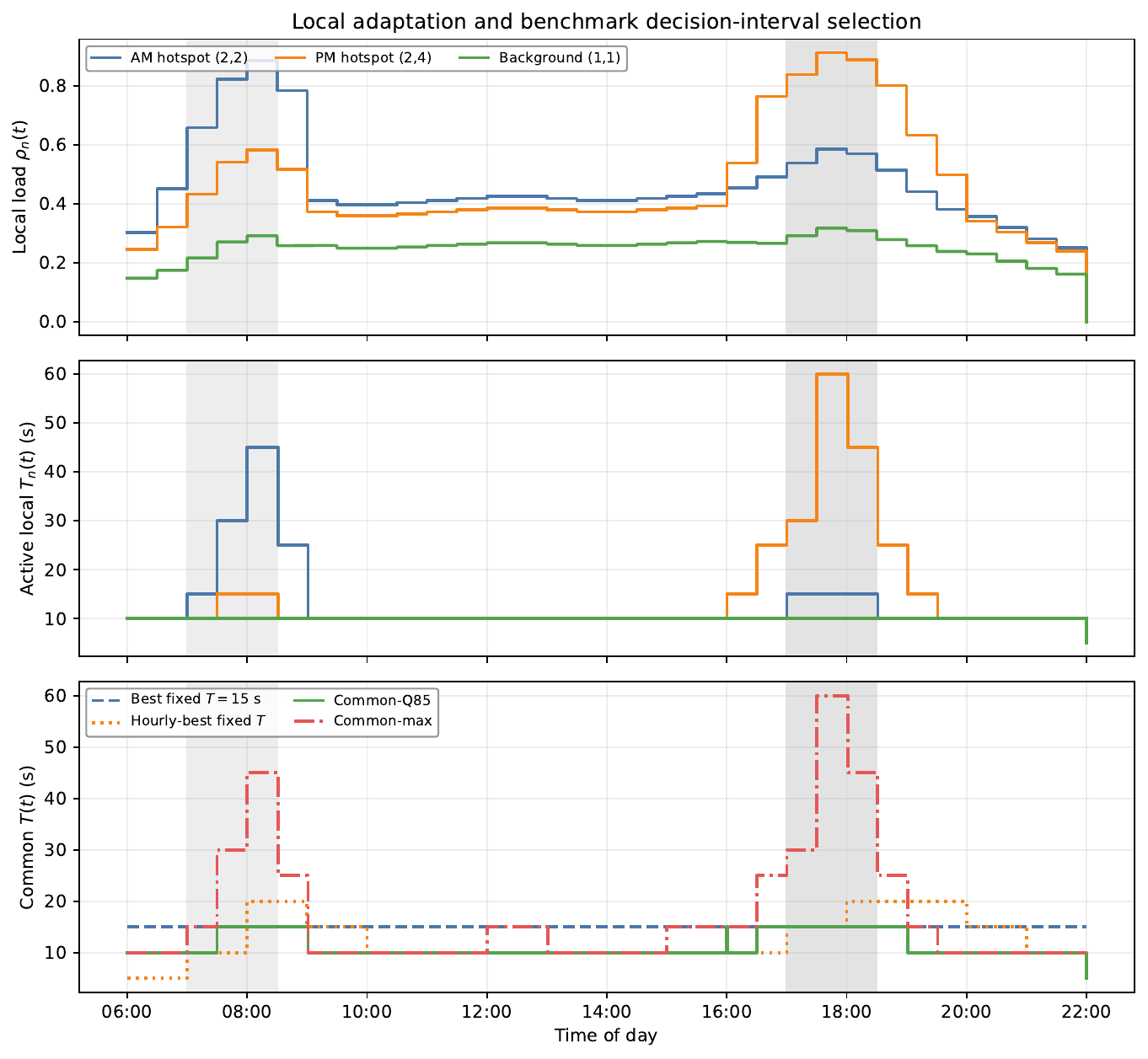}
\caption{Experiment~2: temporal variation of representative local loads and decision intervals. The upper and middle panels show the local load and corresponding heterogeneous VT-MP intervals at representative intersections; the lower panel compares the resulting common timing benchmarks with the best fixed and hourly-best fixed schedules.}
\label{fig:exp2_temporal}
\end{figure*}

The temporal mechanism underlying these differences is shown in Fig.~\ref{fig:exp2_temporal}. The best single fixed interval over the full day is $T=15$~s, while the offline hourly-best benchmark varies between 5 and 20~s, already indicating that the preferred network-wide timescale changes with demand. The local VT-MP recommendations exhibit the same qualitative pattern more directly: hotspot loads rise during the morning and evening peaks and their preferred intervals lengthen, whereas the recommendations contract again as demand subsides.

The common dynamic policies aggregate these local recommendations into a single synchronized network-wide interval, but the aggregation rule strongly affects the resulting behavior. The Q85 rule tracks the increase in network loading without being dictated by the most extreme local condition. The maximum rule, by contrast, is governed by a small number of highly loaded intersections and consequently applies long intervals even to locations that remain lightly loaded. This explains why common temporal adaptation can improve on a fixed interval when the aggregation is moderate, yet still leaves a structural limitation: intersections with different local reserves are forced to share the same decision timescale.

\subsubsection{Spatial heterogeneity}

\begin{figure*}
\centering
\includegraphics[width=0.9\textwidth]{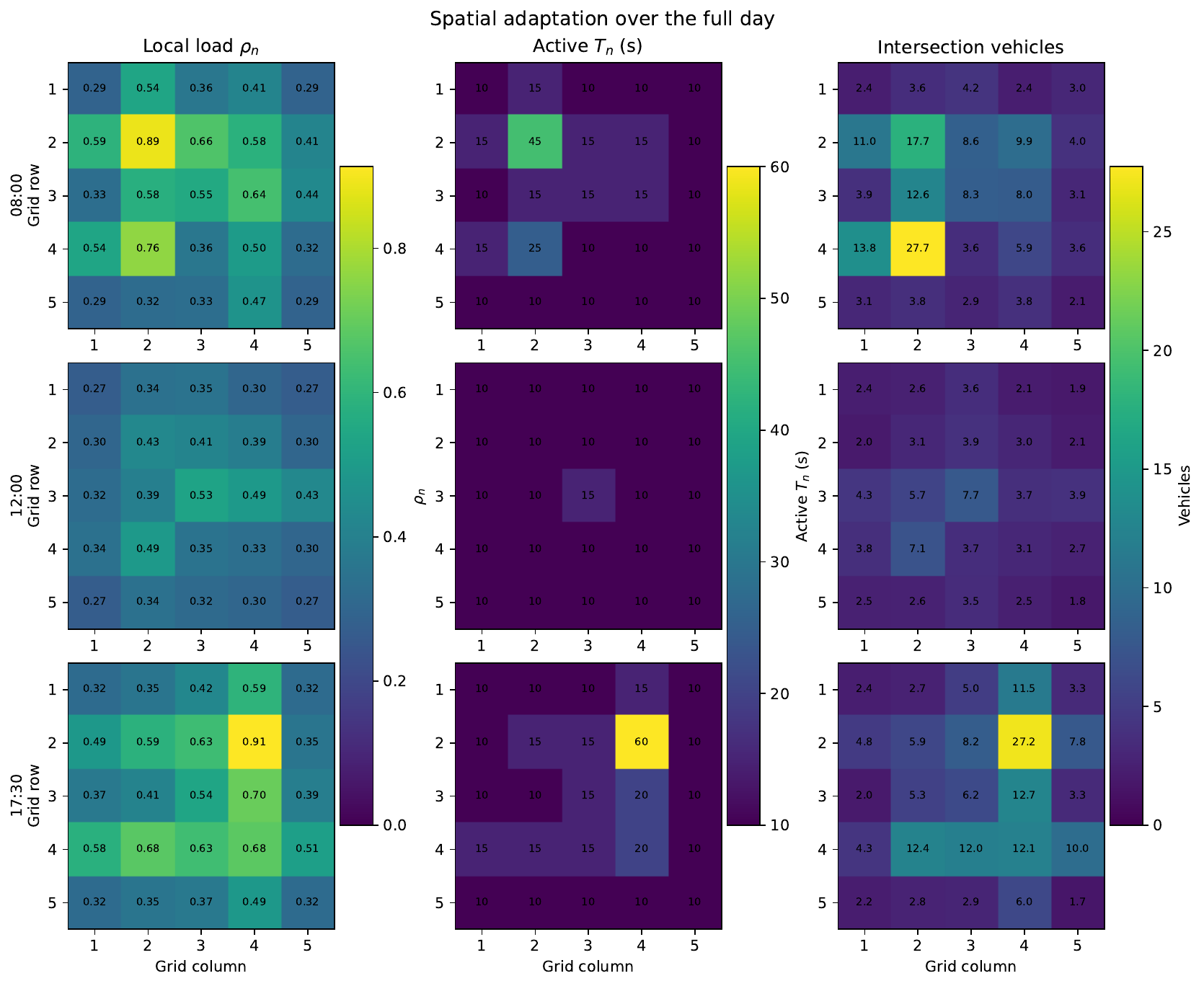}
\caption{Experiment~2: spatial heterogeneity in local load, heterogeneous VT-MP interval, and intersection vehicle accumulation at representative times. Peak-period hotspots receive substantially longer intervals, whereas lightly loaded intersections retain short intervals.}
\label{fig:exp2_spatial}
\end{figure*}

Figure~\ref{fig:exp2_spatial} makes this spatial mismatch explicit. At 08:00, the maximum local load reaches 0.886 and the heterogeneous VT-MP intervals span 10--45~s. At 12:00, the network is substantially more homogeneous: the maximum local load is 0.528 and almost every intersection operates at 10~s. During the evening peak at 17:30, the maximum local load rises to 0.913 and the local intervals span 10--60~s. Thus, the longer recommended intervals are concentrated at a small set of peak-period hotspots rather than being a network-wide requirement.

This spatial differentiation explains why heterogeneous VT-MP achieves a larger gain than the common dynamic policies in Fig.~\ref{fig:exp2_overall}. The controller does not alter the MP pressure calculation; instead, it removes the requirement that every intersection reconsider its pressure decision on the same timescale. Each node can therefore operate at the interval implied by its own local reserve while preserving the common normalized timing principle. This spatial differentiation is consistent with the service-utilization mechanism established in Proposition~\ref{prop:heterogeneous_service_dominance} and with the broader analytical advantage in Proposition~\ref{prop:heterogeneous_bound}. The experiment demonstrates the same principle under network dynamics that closely match the analytical model: temporal adaptation is useful, while the larger gain emerges when the decision timescale is allowed to vary across space as well as time.

\subsection{Experiment 3: Microscopic SUMO network}
\label{subsec:exp_sumo}

Having established network-level effectiveness under the store-and-forward assumptions, the third experiment examines whether the same advantage persists under microscopic finite-capacity traffic. The same $5\times5$ grid is implemented in SUMO, where links have finite length and downstream congestion can restrict upstream discharge. The evaluation covers 07:00--11:00 and is followed by a 30-min clearance period. Demand varies over eight 30-min stages from approximately 5400 to 13,680~veh/h, creating a low-demand period, a two-hour high-demand plateau, and a subsequent decline. Spatial heterogeneity is introduced by four higher-demand OD streams whose fixed routes pass through the central node C2. Three fixed demand/route realizations are used and reused across controllers.

\begin{figure*}
\centering
\includegraphics[width=0.98\textwidth]{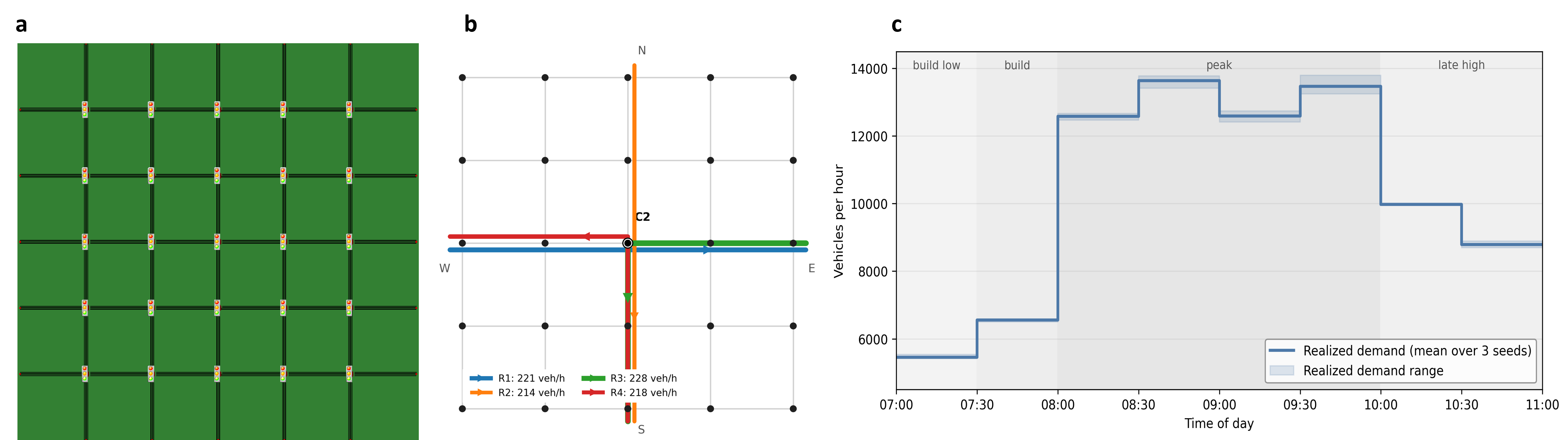}
\caption{Experiment~3 microscopic SUMO setting. Panel~(a) shows the $5\times5$ signalized grid in SUMO. Panel~(b) shows the four strengthened OD routes, all of which traverse the central node C2; arrows indicate travel direction and line width represents the corresponding peak route demand. Panel~(c) shows the realized total demand over the four-hour evaluation period, with the line denoting the mean across the three demand realizations and the shaded band showing their range.}
\label{fig:exp3_scenario}
\end{figure*}

Figure~\ref{fig:exp3_scenario} summarizes the microscopic test setting. Panel~(a) shows the $5\times5$ SUMO network, panel~(b) highlights the four strengthened OD routes and the central bottleneck C2, and panel~(c) shows the realized total demand profile. The four strengthened streams approach C2 from different directions and then continue through the central corridors, intentionally creating a localized high-load region while leaving peripheral intersections more lightly loaded. The realized demand profiles are closely matched across the three realizations, with the high-demand period concentrated between approximately 08:00 and 10:00. This design therefore combines a pronounced temporal demand transition with a persistent spatial concentration of traffic around the center of the network.

Movement arrival rates are estimated from fixed detectors over a 300-s rolling window and converted to local normalized loads. The implementation uses $I=2.5$~s and $\kappa=1.6$, with intervals rounded upward to a 5-s grid. Based on Experiment~1, the load entering the timing formula is capped at $\rho=0.90$, yielding a maximum planned interval of 40~s. For heterogeneous VT-MP, the planned interval is additionally checked against downstream receiving supply and shortened when the neighboring storage constraint becomes active.

The same five benchmark classes used in Experiment~2 are retained: Best fixed, Hourly-best fixed, Common-max, Common-Q85, and Heterogeneous VT-MP. The fixed-$T$ sensitivity again covers 5--60~s in 5-s increments. Traffic performance is evaluated primarily through mean trip delay and network accumulation. To account for vehicles delayed before they can enter a congested source, unfinished demand is defined as
\begin{equation}
N_{\mathrm{unfinished}}(t)
=
N_{\mathrm{due}}(t)-N_{\mathrm{arrived}}(t)
=
N_{\mathrm{overflow}}(t)+N_{\mathrm{in}}(t),
\label{eq:unfinished_vehicles}
\end{equation}
with cumulative unfinished vehicle-hours
\begin{equation}
H_{\mathrm{unfinished}}
=
\int_0^{T_{\mathrm{sim}}}N_{\mathrm{unfinished}}(t)\,\mathrm dt.
\label{eq:unfinished_vehicle_hours}
\end{equation}
For completed trips, delay is the sum of SUMO departure delay and in-network time loss.

\subsubsection{System-level traffic performance}

Table~\ref{tab:exp3_performance} summarizes the system-level traffic performance averaged over the three demand/route realizations. Heterogeneous VT-MP achieves the best performance across all reported measures. Relative to the seed-specific best fixed interval, mean trip delay decreases from 213.3 to 161.5~s, a reduction of 24.3\%, while the mean and peak numbers of vehicles in the network decrease by 14.3\% and 12.3\%, respectively. Cumulative unfinished vehicle-hours are reduced as well. These improvements extend the main finding of Experiment~2 to the microscopic finite-storage network, showing that the advantage of locally adapting the MP decision timescale is retained beyond the store-and-forward setting.

\begin{table*}
\centering
\caption{Experiment~3: microscopic performance averaged over three demand/route realizations.}
\label{tab:exp3_performance}
\begin{tabular}{lrrrr}
\hline
Method & Mean trip delay (s) & Unfinished veh.$\cdot$h & Mean accumulation (veh) & Peak accumulation (veh) \\
\hline
Best fixed & 213.3 & 3944.2 & 976.1 & 1537.0 \\
Hourly-best fixed & 228.3 & 4114.4 & 989.8 & 1693.7 \\
Common-Q85 & 224.9 & 4082.6 & 964.8 & 1652.3 \\
Common-max & 215.6 & 3983.0 & 977.5 & 1661.3 \\
Heterogeneous VT-MP & \textbf{161.5} & \textbf{3357.1} & \textbf{836.1} & \textbf{1347.7} \\
\hline
\end{tabular}
\end{table*}

The common-timescale strategies do not reproduce this improvement. Both the offline hourly schedule and Common-Q85 perform worse than the best fixed benchmark in mean trip delay, while Common-max remains close to it. This contrasts with Experiment~2, where a moderate common dynamic rule still improved the store-and-forward system. The difference is informative: once finite storage becomes active, congestion at one location changes the receiving conditions faced by neighboring intersections, so a single network-wide interval becomes increasingly unable to represent the local control requirement. Temporal adaptation remains relevant, but the microscopic results indicate that its benefit depends on retaining the spatial differentiation of the underlying traffic state.

\begin{figure*}
\centering
\includegraphics[width=0.98\textwidth]{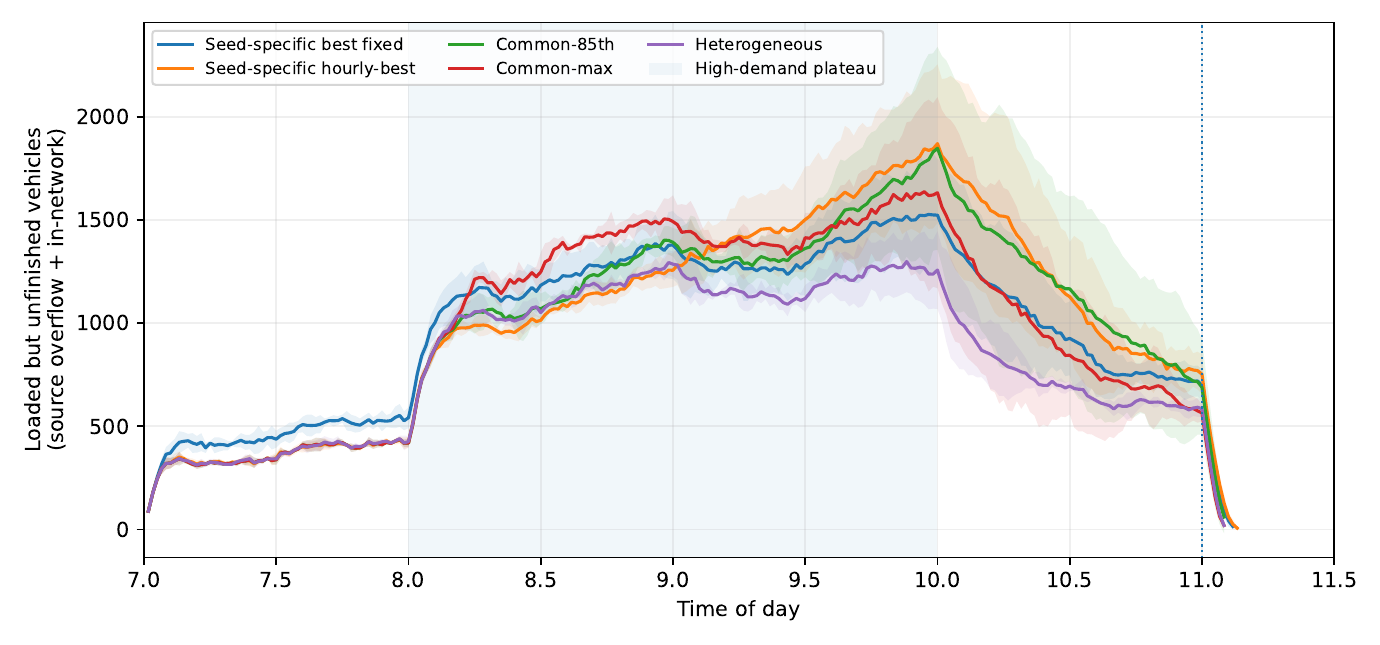}
\caption{Experiment~3: evolution of unfinished demand, including both source overflow and vehicles already inside the network. Lines show the mean across the three demand realizations and shaded bands show between-realization variation. Heterogeneous VT-MP limits accumulation during the high-demand period and accelerates recovery.}
\label{fig:exp3_accumulation}
\end{figure*}

The unfinished-demand evolution in Fig.~\ref{fig:exp3_accumulation} complements the in-network accumulation measures in Table~\ref{tab:exp3_performance} and shows how this system-level advantage develops. The methods remain relatively close during the initial low-demand period, but diverge as the high-demand plateau builds. Heterogeneous VT-MP suppresses the growth of unfinished traffic through the peak and then clears the accumulated demand more rapidly during the recovery period. The lower source-overflow component also confirms that the improvement is not obtained by holding more vehicles outside the modeled network. Taken together with Table~\ref{tab:exp3_performance}, the result shows that heterogeneous VT-MP improves both peak-period congestion containment and post-peak recovery rather than merely shifting one aggregate performance measure. The same qualitative advantage identified in Experiment~2 therefore remains observable in the finite-storage network.

\subsubsection{Temporal evolution of decision timescales}

The temporal results provide a second check on the load--timescale mechanism identified in Experiment~1. The fixed-$T$ sensitivity shows that the preferred common interval is short during the initial low-demand period and becomes substantially longer as the high-demand period develops. Across the three realizations, the best full-horizon fixed interval lies between 25 and 35~s, while the hourly benchmark increases from approximately 15~s early in the simulation to 30--40~s during the most heavily loaded period. Thus, even in the microscopic network, the preferred decision timescale increases with traffic load rather than remaining constant over time.

\begin{figure*}
\centering
\includegraphics[width=0.8\textwidth]{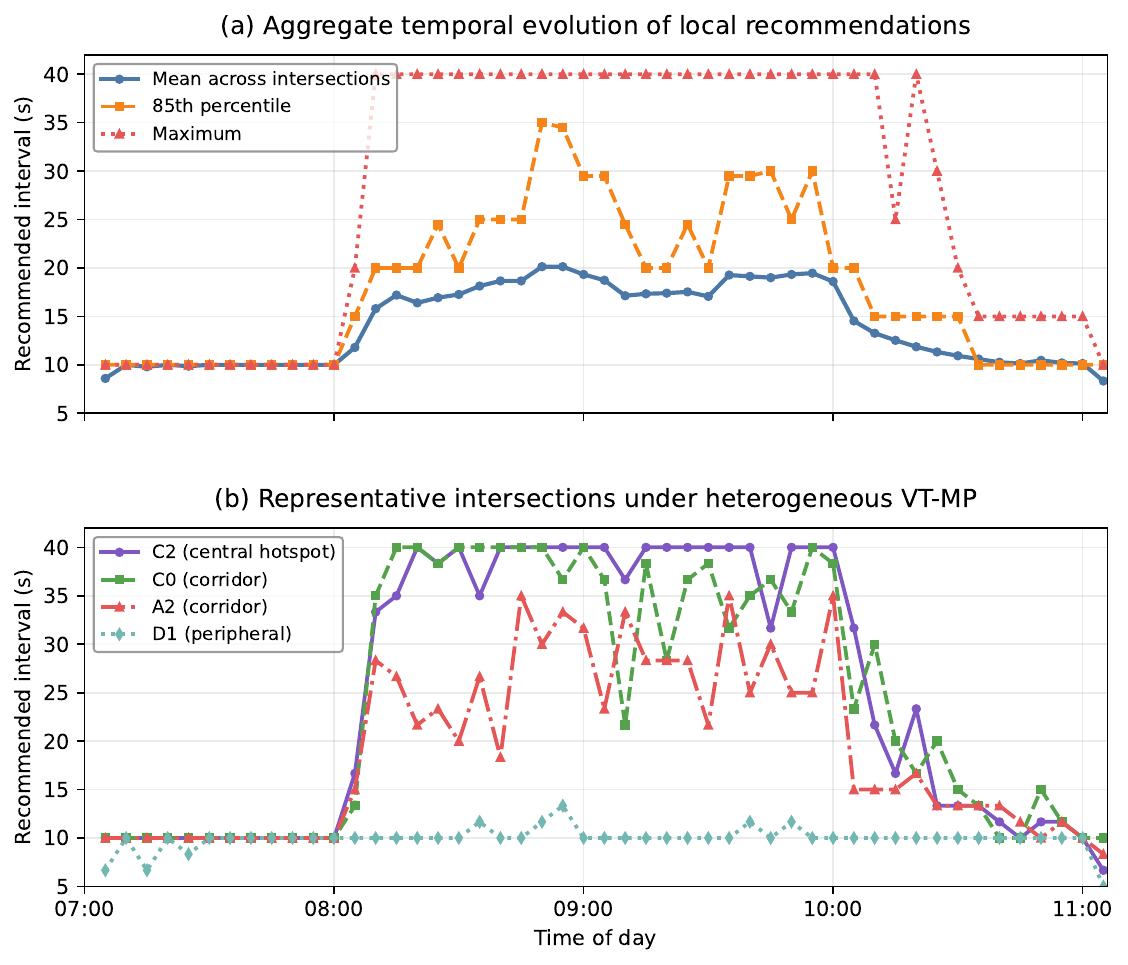}
\caption{Experiment~3: temporal evolution of the detector-based local VT-MP recommendations. Panel~(a) summarizes the distribution across 25 intersections and three demand realizations using the mean, 85th percentile, and maximum local recommendation. Panel~(b) shows the corresponding recommendation histories at representative intersections, illustrating how hotspots, corridors, and peripheral nodes follow different timing trajectories.}
\label{fig:exp3_temporal_T}
\end{figure*}

The online VT-MP recommendations in Fig.~\ref{fig:exp3_temporal_T} reproduce this behavior without selecting intervals from an offline sensitivity analysis. During the initial low-demand period, nearly all intersections remain close to the minimum timing range. As the peak develops, both the average recommendation and the upper tail of the local distribution increase, with the most heavily loaded intersections reaching the 40-s operational cap. The distribution contracts again as demand subsides. A representative comparison is that the network-wide mean recommendation rises from about 10~s in the early period to roughly 19~s near the center of the peak before returning toward 10~s during recovery.

The representative-node trajectories also show that the temporal response is not spatially uniform. C2 and C0 remain at relatively long intervals through much of the peak, A2 follows a more moderate trajectory along the main corridor, whereas the lightly loaded peripheral node D1 remains close to the lower timing range. The operational cap is therefore active only for a limited subset of highly loaded local conditions; most intersections continue to adjust their timescale according to the measured local load. This distinction is important because the temporal increase in the network-average recommendation does not imply that every intersection should slow its update frequency by the same amount.

\subsubsection{Spatial distribution of decision timescales}

\begin{figure*}
\centering
\includegraphics[width=0.8\textwidth]{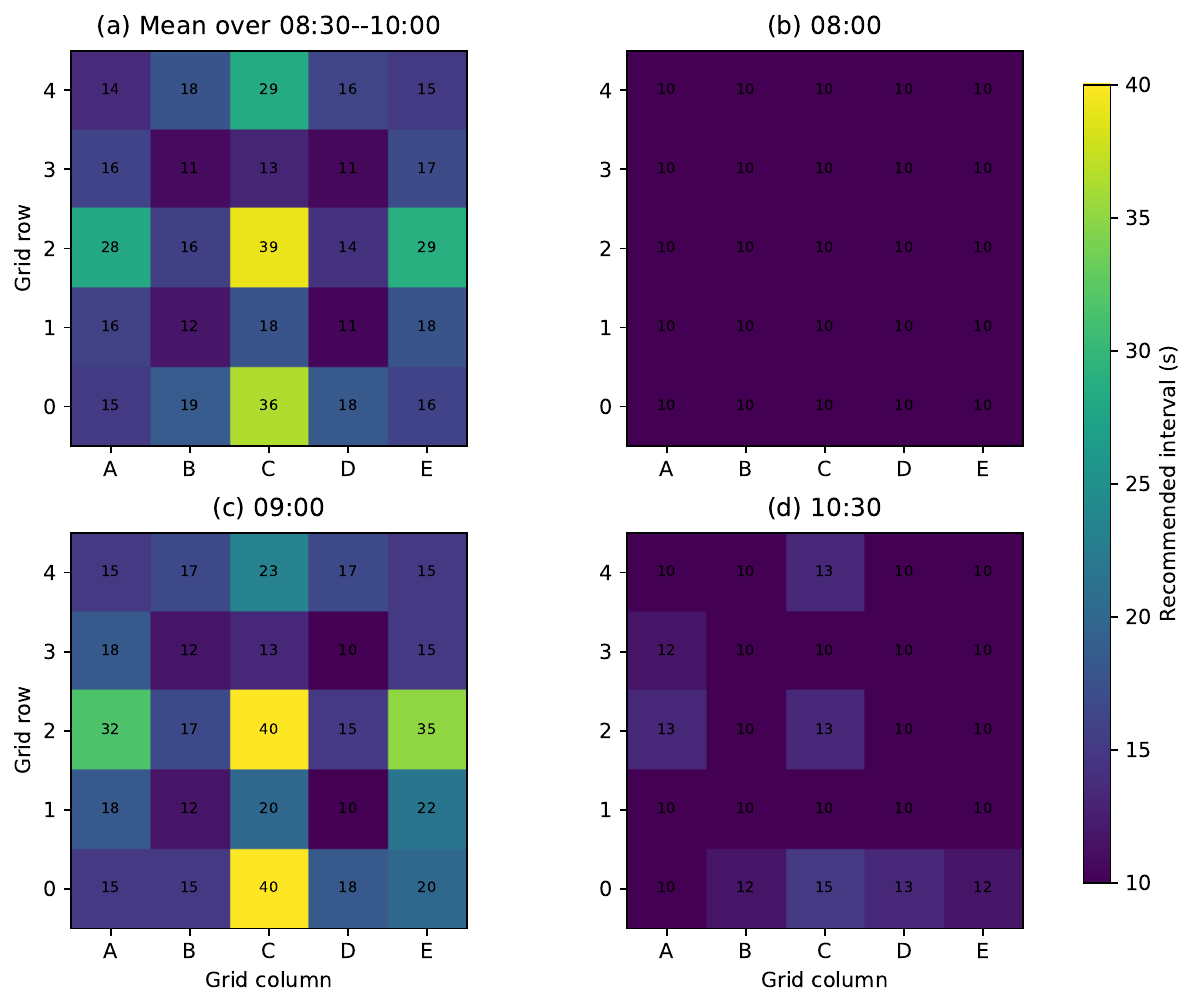}
\caption{Experiment~3: spatial distribution of the local VT-MP recommendation. Panel~(a) shows the mean interval over the high-demand plateau (08:30--10:00), while panels~(b)--(d) show representative time slices at 08:00, 09:00, and 10:30. The longest intervals concentrate along the deliberately strengthened central traffic paths rather than being imposed uniformly across the grid.}
\label{fig:exp3_spatial_T}
\end{figure*}

Figure~\ref{fig:exp3_spatial_T} complements the temporal view by showing where the longer decision intervals occur. Before the high-demand period, the network is nearly uniform and most intersections operate at short intervals. As the central traffic streams intensify, a pronounced spatial structure emerges: the longest recommendations concentrate at C2 and along the strengthened OD corridors, while peripheral intersections remain at substantially shorter intervals. The area requiring longer intervals then contracts as demand declines. Thus, the spatial distribution of the timing recommendation follows the spatial distribution of traffic load rather than a network-wide change in operating scale.

This spatial pattern is strongly aligned with the estimated local operating condition. Across the 25 intersections, the mean peak-period recommendation has a Pearson correlation of 0.97 with the corresponding detector-based load. The most heavily loaded central nodes consequently operate near the upper timing range, whereas lightly loaded peripheral nodes retain short intervals even while the network as a whole is congested. This spatial pattern reproduces, under finite-capacity microscopic traffic, the mechanism observed in Experiment~2: the performance gain does not come from choosing a different network-wide average interval, but from allowing the decision timescale to expand only where the local service reserve becomes small and to remain short elsewhere.

Overall, Experiment~3 demonstrates that the network-level effectiveness observed in Experiment~2 carries to microscopic finite-capacity traffic. The temporal and spatial timing patterns remain consistent with the load-dependent mechanism, while heterogeneous VT-MP continues to outperform both fixed and common dynamic timing strategies under online load estimation, finite storage, spillback, and receiving-supply limitations.

\subsection{Discussion and limitations}
\label{subsec:discussion_limitations}

The three experiments form a consistent validation sequence. The single-intersection mechanism study first validates the theoretical load--timescale relationship by showing that the preferred decision interval grows with load and follows the characteristic scaling $T=\kappa I/(1-\rho)$ over the practically relevant subcritical range. The store-and-forward experiment then evaluates the effectiveness of the resulting spatiotemporally heterogeneous timing policy in a network that closely matches the analytical assumptions, showing that temporal adaptation improves on a single fixed interval and that a larger gain is obtained when intersections operate at different local timescales. The microscopic experiment extends the effectiveness test to finite-capacity traffic, where local load is estimated online and downstream storage and spillback physically couple neighboring intersections.

The comparison between Experiments~2 and~3 also demonstrates the practical applicability of the analytical timing mechanism. Store-and-forward models have long provided a tractable basis for network-wide traffic signal control \citep{diakaki2002tuc,aboudolas2009store}, while finite-storage and downstream-capacity effects are known to become important as congestion propagates through the network \citep{gregoire2015capacity,li2019position,hao2020downstream}. In this comparison, the infinite-storage analysis isolates the trade-off between switching efficiency and decision staleness and provides the normalized load-dependent timing scale. Experiment~2 verifies this mechanism under conditions that closely follow those assumptions, whereas Experiment~3 shows that the same timing principle remains useful when the physical receiving space of downstream links becomes an additional constraint. In that case, the analytical rule provides the load-dependent planning timescale, while the receiving-supply correction determines how much of the planned interval can safely be executed under the current finite-storage state.

A further limitation arises close to the capacity boundary. Because $T_c=I/(1-\rho)$ diverges as $\rho\rightarrow1$, small errors in a near-unit load estimate can generate very large changes in the recommended interval. Experiment~1 further shows that finite-storage spillback can eliminate the microscopic feasible interval before the ideal point-queue boundary is reached. The $\rho=0.90$ cap used in Experiment~3 is consequently tied to the empirically validated range of the single-intersection calibration. Extending VT-MP deeper into oversaturated traffic would require an explicit finite-storage and recovery model in the theory itself.

Finally, both network experiments use a structured $5\times5$ grid so that the temporal and spatial mechanisms can be observed directly. The present study therefore focuses on developing and validating the decision-timescale theory and its network-level implications rather than exhaustively testing irregular urban networks. For the same reason, the experiments assume that the traffic states and model parameters required by the controller, such as local demand, turning proportions, and service characteristics, can be observed or estimated with sufficient accuracy, so that the effect of the proposed timing mechanism can be isolated from estimation uncertainty. This assumption does not imply that the controller requires exact parameter values. The normalized policy admits a range of timing factors, $1<\kappa\le2$, and the single-intersection results show that favorable performance is typically obtained over a range of decision intervals rather than at a single isolated value, particularly under lower traffic loads where the mapping from $\rho$ to $T$ is less sensitive. In addition, several key quantities required by the controller are comparatively amenable to field estimation: traffic flow can be obtained from conventional or video-based detectors, while the effective switching loss can be calibrated from field observations or vehicle-trajectory data, including aerial trajectory measurements when available. Natural extensions include relaxing the accurate-information assumption by explicitly considering uncertainty in both traffic-state measurements and parameter estimation, together with heterogeneous phase structures and larger-scale coordinated networks.

\section{Conclusions}
\label{sec:conclusion}

This paper examined a dimension of Max-Pressure control that is usually fixed in advance: the timescale at which the phase decision is reconsidered. By explicitly accounting for the effective switching loss $I$ over a decision interval $T$, we derived a switching-adjusted capacity guarantee that scales with the productive fraction $1-I/T$. This yields the load-dependent critical timescale
$T_c(\rho)=I/(1-\rho)$,
which marks the boundary at which the switching-adjusted service reserve becomes positive. A finite-frame Lyapunov analysis then gives a queue-bound scaling proportional to $T^2/[(1-\rho)T-I]$, whose characteristic minimizer is $2T_c$. Together, these results identify the performance-oriented range $T_c<T\le2T_c$ and establish the central scaling $T\propto I/(1-\rho)$.

Building on this relation, we proposed Variable-Timescale Max-Pressure control, in which each intersection uses a local decision interval
$T_n=\kappa I_n/(1-\rho_n)$.
The formulation preserves the conventional MP phase-selection rule while allowing the update frequency to vary across both time and space. Under uniformly admissible local loads and bounded local intervals, the resulting asynchronous heterogeneous controller retains strong stability, and this stability result extends to sufficiently persistent piecewise-stationary demand. The analysis further shows that heterogeneous timing can yield strictly greater realized service when a common interval outlasts the useful service duration of a local phase, while removing the common-timescale constraint also improves the best attainable analytical queue-bound surrogate. For microscopic networks, a downstream receiving-supply correction treats the theoretical interval as a planned timescale and shortens it when finite downstream storage becomes restrictive.

The experiments provide two complementary forms of evidence. First, the single-intersection mechanism study validates the theory: the preferred interval grows systematically with traffic load and follows the derived load-dependent characteristic scales, while also yielding $\kappa\approx1.6$ for the subsequent network tests. Second, the two network experiments establish effectiveness. In the store-and-forward network, which closely matches the analytical assumptions, heterogeneous VT-MP reduces total queueing time by 31.0\% relative to the best fixed interval and reveals clear temporal and spatial variation in the preferred timescale. The microscopic SUMO experiment then shows that this advantage persists in a finite-capacity network with detector-based state estimation, downstream receiving constraints, and spillback: mean trip delay is reduced by 24.3\% and peak network accumulation by 12.3\% relative to the seed-specific best fixed benchmark. Together, the results link the theoretical load--timescale mechanism to consistent network-level gains under both theory-consistent and microscopic traffic dynamics.

The main practical boundary of the present framework occurs near capacity. As $\rho$ approaches one, the analytical timescale becomes increasingly sensitive to the load estimate, while finite-storage spillback can become critical before the ideal point-queue capacity boundary is reached. The microscopic implementation therefore limits the timing recommendation to the load range validated by the single-intersection experiment and separately enforces downstream receiving feasibility. Future work should extend the theory to explicitly finite-storage and oversaturated traffic, investigate uncertainty in local load and turning estimates, and evaluate the approach on larger irregular urban networks and multimodal control settings. The central conclusion is that the MP decision timescale is not merely an implementation parameter: it is a traffic-state-dependent control variable that can be designed jointly with the phase decision.

\appendix

\section{Detailed derivation of the T-frame drift bound}
\label{app:Tslot_drift}

This appendix supplies the bounded-error steps used in Lemma~\ref{lem:Tslot_drift}. From Eq.~\eqref{eq:queue_dynamics}, the standard quadratic inequality gives
\begin{align}
V(\mathbf{Q}(t+1))-V(\mathbf{Q}(t))
\le\;&
B_0
+
\sum_{(a,b)\in\mathcal{M}}
Q_{ab}(t)A^{\mathrm{ext}}_{ab}(t)
-
\sum_{(a,b)\in\mathcal{M}}
w_{ab}(t)D_{ab}(t),
\label{eq:appendix_one_slot}
\end{align}
where the internal routed-flow terms have been reindexed into the movement pressures and $B_0<\infty$ follows from bounded arrivals and services.

Define the unused potential service
\[
U_{ab}(t)
=
S_{ab}(t)-D_{ab}(t)\ge0.
\]
Using $D_{ab}=S_{ab}-U_{ab}$,
\begin{equation}
-\sum_{(a,b)}w_{ab}D_{ab}
=
-\sum_{(a,b)}w_{ab}S_{ab}
+
\sum_{(a,b)}w_{ab}U_{ab}.
\label{eq:unused_decomposition}
\end{equation}
Let $S_{\max}=\max_{(a,b),p}\mu_{ab}(p)$. Whenever $U_{ab}(t)>0$, Eq.~\eqref{eq:departure_constraint} implies $Q_{ab}(t)<S_{ab}(t)\le S_{\max}$. Since $w_{ab}(t)\le Q_{ab}(t)$,
\[
w_{ab}^{+}(t)U_{ab}(t)
\le
S_{\max}^2.
\]
The negative part of $w_{ab}U_{ab}$ only decreases the drift. Hence,
\begin{equation}
\sum_{(a,b)\in\mathcal{M}}
w_{ab}(t)U_{ab}(t)
\le
|\mathcal{M}|S_{\max}^2
\equiv
B_u,
\label{eq:unused_service_constant}
\end{equation}
which establishes the bounded unused-service correction used in the proof of Lemma~\ref{lem:Tslot_drift}.

Now sum Eq.~\eqref{eq:appendix_one_slot} over $\tau=0,\ldots,T-1$. Because arrivals and services are bounded, there exists $G_Q>0$ such that
\begin{equation}
\|\mathbf{Q}(t+\tau)-\mathbf{Q}(t)\|_1
\le
G_Q\tau.
\label{eq:queue_increment_bound}
\end{equation}
Movement pressure is a fixed linear transformation of the queue vector under the fixed turning proportions. Therefore, there exists $G_w>0$ such that
\begin{equation}
\|\mathbf{w}(t+\tau)-\mathbf{w}(t)\|_1
\le
G_w\tau.
\label{eq:pressure_increment_bound}
\end{equation}
The difference between the drift evaluated with the evolving queues and pressures and the corresponding frame-initial quantities is consequently bounded by
\begin{equation}
\sum_{\tau=0}^{T-1}
\left(
C_QG_Q\tau+C_wG_w\tau
\right)
\le
B_1T^2
\label{eq:frame_state_error}
\end{equation}
for finite constants $C_Q,C_w,B_1$ determined by the arrival and service bounds.

It remains to bound the frame-initial term. Let $\phi_T=\phi(T)=1-I/T$. Since $\boldsymbol{\lambda}\in\rho\mathcal C$ and $\rho<\phi_T$, the standard MP comparison construction provides a pressure-domain reserve proportional to $\phi_T-\rho$. Hence, for some constant $\chi_r>0$ independent of $T$ and $\mathbf Q(t)$,
\begin{align}
&\mathbb{E}\Bigg[
\sum_{\tau=0}^{T-1}
\Bigg(
\sum_{(a,b)\in\mathcal{M}}
Q_{ab}(t)A^{\mathrm{ext}}_{ab}(t+\tau)
-
\sum_{(a,b)\in\mathcal{M}}
w_{ab}(t)S_{ab}(t+\tau)
\Bigg)
\;\Bigg|\;
\mathcal{F}_t
\Bigg]
\le
-\chi_rT(\phi_T-\rho)
\|\mathbf w^{+}(t)\|_1.
\label{eq:appendix_pressure_comparison}
\end{align}
To convert the pressure reserve to a queue reserve, define the nonnegative routing operator $\mathsf R$ by
\[
(\mathsf R\mathbf Q)_{ab}
=
\sum_{c:(b,c)\in\mathcal M}r_{bc}Q_{bc}.
\]
Equation~\eqref{eq:movement_pressure} gives $\mathbf w=(\operatorname{Id}-\mathsf R)\mathbf Q$. Open routing implies that $(\operatorname{Id}-\mathsf R)^{-1}$ exists and is nonnegative. Writing $\mathbf w=\mathbf w^{+}-\mathbf w^{-}$ therefore gives
\[
\mathbf Q
=(\operatorname{Id}-\mathsf R)^{-1}\mathbf w
\le
(\operatorname{Id}-\mathsf R)^{-1}\mathbf w^{+}.
\]
Consequently,
\begin{equation}
\|\mathbf w^{+}(t)\|_1
\ge
\gamma_R\|\mathbf Q(t)\|_1,
\qquad
\gamma_R
=
\left\|(\operatorname{Id}-\mathsf R)^{-1}\right\|_1^{-1}>0.
\label{eq:appendix_pressure_queue_norm}
\end{equation}
Combining Eqs.~\eqref{eq:appendix_pressure_comparison} and \eqref{eq:appendix_pressure_queue_norm}, with $\chi=\chi_r\gamma_R$, yields
\begin{equation}
-\chi_rT(\phi_T-\rho)\|\mathbf w^{+}(t)\|_1
\le
-\chi\big[(1-\rho)T-I\big]\|\mathbf Q(t)\|_1.
\label{eq:appendix_comparison}
\end{equation}
The pressure-maximizing phase selected at the frame start is used in the productive portion of the frame, so its frame-initial pressure-weighted potential service is no smaller than that of the feasible comparison allocation.

Combining Eqs.~\eqref{eq:unused_service_constant}, \eqref{eq:frame_state_error}, and \eqref{eq:appendix_comparison} with the summed one-slot drift gives
\[
\mathbb{E}
\left[
V(\mathbf{Q}(t+T))-V(\mathbf{Q}(t))
\mid\mathcal{F}_t
\right]
\le
(B_0+B_u)T+B_1T^2
-\chi\Delta_T(\rho)\|\mathbf{Q}(t)\|_1.
\]
For $T\ge1$, the first two bounded terms can be absorbed into $BT^2$ for a $T$-independent constant $B$, which proves Eq.~\eqref{eq:Tslot_drift}.

\section{Detailed proof of asynchronous VT-MP stability}
\label{app:async_stability}

This appendix supplies the complete-frame and boundary-fragment bounds used in Theorem~\ref{thm:async_stability}. Under the fixed mean operating condition, the effective movement flow $\boldsymbol\nu$ satisfies Eq.~\eqref{eq:effective_flow_balance}; therefore, Eq.~\eqref{eq:flow_pressure_identity} converts the exogenous-arrival part of the Lyapunov drift into a pressure-weighted offered-flow term.

For node $n$, let $\phi_n=1-I_n/T_n$ and $\sigma_n=\phi_n-\rho_n$. The theorem assumptions give $\sigma_n\ge\sigma_{\min}>0$. Since $\boldsymbol\nu_n\in\rho_n\mathcal S_n$, there exists $\mathbf s_n^{0}\in\mathcal S_n$ such that $\boldsymbol\nu_n\le\rho_n\mathbf s_n^{0}$. Fix the strictly positive vector $\mathbf d_n\in\mathcal S_n$ used in the proof of Theorem~\ref{thm:async_stability}. By convexity,
\begin{equation}
\rho_n\mathbf s_n^{0}
+
(\phi_n-\rho_n)\mathbf d_n
\in
\phi_n\mathcal S_n.
\label{eq:app_local_convex_reserve}
\end{equation}
Because $\mathcal S_n$ is downward closed, it contains a vector $\bar{\mathbf s}_n\ge\boldsymbol\nu_n+(\phi_n-\rho_n)\mathbf d_n$. For any pressure vector $\mathbf w_n$, set the components of $\bar{\mathbf s}_n$ corresponding to nonpositive pressures to zero. The resulting vector remains in $\phi_n\mathcal S_n$, and
\begin{equation}
\mathbf w_n^{\top}\boldsymbol\nu_n
-
\mathbf w_n^{\top}\bar{\mathbf s}_n^{\,\mathbf w}
\le
-\sigma_n d_{\min}\|\mathbf w_n^{+}\|_1.
\label{eq:app_local_pressure_reserve}
\end{equation}
Summing over nodes and using Eq.~\eqref{eq:pressure_queue_norm} gives the queue-dependent reserve $\eta\|\mathbf Q\|_1$ with $\eta=\sigma_{\min}d_{\min}\gamma_R>0$.

Now let $\tau_n(s)$ denote the update epoch associated with the phase retained by node $n$ at slot $s$. Since $T_n\le T_{\max}$,
\[
0\le s-\tau_n(s)\le T_{\max}.
\]
Bounded one-slot queue increments imply
\[
\|\mathbf Q(s)-\mathbf Q(\tau_n(s))\|_1
\le
G_QT_{\max},
\]
and linearity of phase pressure therefore gives, for some $L_P<\infty$,
\[
\left|
P_{n,p}(s)-P_{n,p}(\tau_n(s))
\right|
\le
L_PT_{\max}
\qquad
\forall p\in\mathcal P_n.
\]
If $p_n(s)$ is the phase selected at $\tau_n(s)$, then on every productive slot
\begin{align}
P_{n,p_n(s)}(s)
&\ge
P_{n,p_n(s)}(\tau_n(s))-L_PT_{\max}
=
\max_{p\in\mathcal P_n}P_{n,p}(\tau_n(s))-L_PT_{\max}
\ge
\max_{p\in\mathcal P_n}P_{n,p}(s)-2L_PT_{\max}.
\label{eq:app_approximate_mp}
\end{align}
Thus asynchronous retention creates an additive pressure error independent of the queue magnitude.

Fix $H>2T_{\max}$. At each node, remove the local frame intersecting the left boundary of $[t,t+H)$ and the local frame intersecting the right boundary. The remaining complete local frames cover at least $H-2T_{\max}$ slots. Over each such frame, the productive duration is $T_n-I_n=\phi_nT_n$. Since the retained phase maximizes pressure at its frame start, its frame-initial pressure-weighted productive service is no smaller than that of any average comparison vector in $\phi_n\mathcal S_n$. Equations~\eqref{eq:app_local_pressure_reserve} and \eqref{eq:app_approximate_mp} therefore give a negative complete-frame drift contribution proportional to the total complete-frame duration.

The queue state at any $s\in[t,t+H)$ satisfies
\begin{equation}
\|\mathbf Q(s)-\mathbf Q(t)\|_1
\le
G_QH.
\label{eq:app_horizon_queue_difference}
\end{equation}
Replacing the frame-start queue and pressure vectors by their values at $t$ consequently creates an $O(H^2)$ state-independent error. Collecting this error, the quadratic Lyapunov remainders, unused-service corrections, and the bounded additive loss in Eq.~\eqref{eq:app_approximate_mp} gives
\begin{equation}
\mathbb E
\left[
\Delta_H^{\mathrm{comp}}V(t)
\mid\mathcal F_t
\right]
\le
B_H^{(1)}
-
 \chi_0\eta(H-2T_{\max})
\|\mathbf Q(t)\|_1
\label{eq:app_complete_portion}
\end{equation}
for finite $B_H^{(1)}$ and $\chi_0>0$.

The two boundary fragments at each node have total duration at most $2T_{\max}$. Bounded arrivals and services imply
\begin{equation}
\mathbb E
\left[
\Delta_H^{\partial}V(t)
\mid\mathcal F_t
\right]
\le
B_H^{(2)}
+
 \chi_{\partial}T_{\max}\|\mathbf Q(t)\|_1
\label{eq:app_boundary_portion}
\end{equation}
for finite $B_H^{(2)}$ and $\chi_{\partial}>0$. Adding Eqs.~\eqref{eq:app_complete_portion} and \eqref{eq:app_boundary_portion}, with $B_H=B_H^{(1)}+B_H^{(2)}$, gives
\begin{equation}
\mathbb E
\left[
V(\mathbf Q(t+H))-V(\mathbf Q(t))
\mid\mathcal F_t
\right]
\le
B_H
-
 \chi_0\eta(H-2T_{\max})\|\mathbf Q(t)\|_1
+
 \chi_{\partial}T_{\max}\|\mathbf Q(t)\|_1.
\label{eq:async_drift_before_H}
\end{equation}
Because $\eta>0$, a finite $H$ can be chosen so that $\delta_H=\chi_0\eta(H-2T_{\max})-\chi_{\partial}T_{\max}>0$, proving the drift condition in Eq.~\eqref{eq:general_H_drift}.

Finally, summing the $H$-step drift over $m=0,\ldots,M-1$ gives $\delta_H\sum_{m=0}^{M-1}\mathbb E[\|\mathbf Q(mH)\|_1]\le MB_H+\mathbb E[V(\mathbf Q(0))]$. After division by $M\delta_H$ and passage to the limit, the sampled-epoch average is finite. For $0\le\tau<H$, Eq.~\eqref{eq:app_horizon_queue_difference} gives
\[
\|\mathbf Q(mH+\tau)\|_1
\le
\|\mathbf Q(mH)\|_1+G_Q\tau,
\]
so the all-slot time average is finite as well, establishing the strong-stability condition in Eq.~\eqref{eq:strong_stability_definition}.

\subsection{Piecewise-stationary extension}

For Corollary~\ref{cor:piecewise_stationary}, the uniform bound $\rho_n^{(m)}\le\bar\rho<1$ makes $T_{\max}$, $\sigma_{\min}$, $\eta$, and the constants in the stale-pressure bound uniform over all operating periods. After the mean arrival rate changes, a node may continue executing at most one interval selected under the preceding operating condition; therefore all nodes adopt the new condition within at most $T_{\max}$ slots. The remainder of the operating period can then be partitioned into complete $H$-windows, with at most one shorter interval left at the end. These short intervals have uniformly bounded total duration, yielding a boundary drift of the form $\mathbb E[\Delta V_{\partial}\mid\mathcal F_s]\le B_{\partial}+\xi_{\partial}\|\mathbf Q(s)\|_1$.

Let $s$ denote the beginning of the adjustment interval and let $s_j$ denote the start of the $j$th subsequent complete $H$-window. If $J^\star$ complete windows are contained in the same operating period, then bounded queue increments give
\[
\|\mathbf Q(s_j)\|_1
\ge
\|\mathbf Q(s)\|_1
-
G_Q\bigl(T_{\max}+J^\star H+H\bigr),
\qquad
j=0,\ldots,J^\star-1.
\]
Summing this inequality gives $\sum_{j=0}^{J^\star-1}\|\mathbf Q(s_j)\|_1\ge J^\star\|\mathbf Q(s)\|_1-C^\star$ with
\[
C^\star
=
J^\star G_Q
\bigl(T_{\max}+J^\star H+H\bigr).
\]
The $J^\star$ interior drift inequalities contribute
\[
J^\star B_H
-
\delta_H
\sum_{j=0}^{J^\star-1}
\|\mathbf Q(s_j)\|_1.
\]
Choose $J^\star$ so that $J^\star\delta_H>\xi_{\partial}$. Adding the boundary contribution then gives a block drift
\begin{equation}
\mathbb E
\left[
V(\mathbf Q(s+L^\star))-V(\mathbf Q(s))
\mid\mathcal F_s
\right]
\le
B^\star
-
\delta^\star\|\mathbf Q(s)\|_1,
\qquad
\delta^\star=J^\star\delta_H-\xi_{\partial}>0.
\label{eq:piecewise_block_drift}
\end{equation}
It is sufficient to choose
\[
L^\star
>
T_{\max}
+
J^\star H
+
H,
\]
so that every operating period contains the initial adjustment interval, $J^\star$ complete $H$-windows, and any remaining interval shorter than $H$. The resulting block drift has a common positive coefficient $\delta^\star$, and the standard telescoping argument establishes strong stability across the sequence of operating periods.

\section*{Declaration of generative AI and AI-assisted technologies in the writing process}

During the preparation of this work, the authors used ChatGPT in order to improve the language and assist with programming for experiments. After using this tool/service, the authors reviewed and edited the content as needed and take full responsibility for the content of the publication.

\bibliographystyle{plainnat}
\bibliography{references}

\end{document}